\documentclass[10pt]{article}
\usepackage[preprint]{tmlr}

\usepackage{amsmath,amsfonts,bm}

\def\eqref#1{equation~\ref{#1}}

\def\1{\bm{1}}

\def\vf{{\bm{f}}}

\def\vk{{\bm{k}}}

\def\vr{{\bm{r}}}

\def\vx{{\bm{x}}}
\def\vy{{\bm{y}}}

\def\mI{{\bm{I}}}

\def\mK{{\bm{K}}}

\DeclareMathAlphabet{\mathsfit}{\encodingdefault}{\sfdefault}{m}{sl}
\SetMathAlphabet{\mathsfit}{bold}{\encodingdefault}{\sfdefault}{bx}{n}

\def\gG{{\mathcal{G}}}

\newcommand{\R}{\mathbb{R}}

\newcommand{\Var}{\mathrm{Var}}

\DeclareMathOperator*{\argmax}{arg\,max}

\usepackage{hyperref}
\hypersetup{hidelinks}
\usepackage{url}
\usepackage{booktabs}
\usepackage{multirow}
\usepackage{amsmath,amssymb,amsthm}
\usepackage{mathtools}
\usepackage{xcolor}
\usepackage{enumitem}
\usepackage{algorithm}
\usepackage{algpseudocode}
\usepackage{subcaption}
\usepackage{microtype}

\newcommand{\PP}{\mathbb{P}}
\newcommand{\kvec}{\vk}             
\newcommand{\resvec}{\vr}           
\DeclareMathOperator{\logdet}{log\,det}
\newcommand{\norm}[1]{\left\|#1\right\|}

\newtheorem{theorem}{Theorem}
\newtheorem{lemma}[theorem]{Lemma}

\newtheorem{remark}{Remark}
\newtheorem{assumption}{Assumption}

\title{\textbf{InfoDPP-PAC}: Principled Patch Selection for Whole Slide Image
  Analysis}

\author{\name Prateek Mittal \email prateek.mittal@feaih.org \\
      \addr Vision Exploration and Data Analytics (VEDAs) Lab \\
      Motilal Nehru National Institute of Technology Allahabad, Prayagraj 211004, India
      \AND
      \name Ayush Srivastava \\
      \addr Vision Exploration and Data Analytics (VEDAs) Lab \\
      Motilal Nehru National Institute of Technology Allahabad, Prayagraj 211004, India
      \AND
      \name Joohi Chauhan  \email joohi@mnnit.ac.in \\
      \addr (Corresponding Author) \\Vision Exploration and Data Analytics (VEDAs) Lab\\
      Motilal Nehru National Institute of Technology Allahabad, Prayagraj 211004, India
      }

\begin{document}
\maketitle

\begin{abstract}
Whole-slide image (WSI) analysis is limited by a familiar mismatch: each slide
contains tens of thousands of candidate tissue patches, while supervision is
usually available only at slide level. Existing bag-construction strategies tend
to address only one side of this problem. Uniform extraction and handcrafted
heuristics do not control redundancy, attention-based multiple-instance models
couple patch importance to a particular downstream classifier, and coreset
methods optimise embedding-space coverage without modelling task-relevant patch
quality. We introduce \textbf{InfoDPP-PAC}, a principled patch-selection
framework that combines teacher-seeded Gaussian process relevance modelling,
determinantal log-determinant diversity, submodular greedy optimisation, and a
concentration-based adaptive stopping rule. The main theoretical result shows
that the log-determinant diversity term used in DPP-style selection is the
Gaussian process mutual information between a selected subset and the latent
relevance function. This yields a monotone submodular objective with standard
greedy approximation guarantees at fixed budgets. We further derive a PAC-style
certificate for residual information gain, allowing the number of retained
patches to vary by slide rather than being fixed a priori. The empirical study
is deliberately scoped to \emph{selection-quality validation}: it evaluates
whether the selected subset is diverse, spatially and morphologically covering,
non-redundant, and enriched for the teacher-derived relevance signal. It does
not claim end-to-end diagnostic improvement after retraining a downstream MIL
model. On 202 HISTAI gastrointestinal whole-slide images, the adaptive rule
uses 83.7\% fewer patches on average than a fixed full budget while retaining
97.9\% of full-budget composite selection quality. At a matched budget,
InfoDPP-PAC achieves the highest mean teacher-derived relevance score among
fourteen baselines, with diversity and composite scores close to the strongest
coreset methods. The results support InfoDPP-PAC as a controlled
quality-diversity-cardinality selection framework, rather than as a downstream
clinical predictor.
\end{abstract}

\section{Introduction}
\label{sec:intro}

Multiple Instance Learning (MIL) has emerged as the dominant paradigm for
Whole Slide Image analysis, treating each WSI as a bag of patches and inferring
slide-level labels without dense spatial annotation
\citep{campanella2019clinical,ilse2018attention,lu2021data}.
A pre-processing step largely orthogonal to the choice of MIL aggregator,
yet one that determines the computational and evidential content presented to
downstream models, is \emph{patch selection}: which subset of the $10^4$--$10^5$ candidate tissue patches to
retain before training or inference.
The canonical choice remains uniform grid extraction
\citep{campanella2019clinical}, which processes every patch above a tissue
threshold and thereby wastes computation on redundant morphological regions.

The literature has proposed several more principled alternatives.
Heuristic scoring (entropy, edge density, texture energy) prioritises patches
with high handcrafted complexity, but provides no information-theoretic
optimality guarantee and no explicit mechanism to penalise redundancy.
Attention-based MIL architectures \citep{ilse2018attention,lu2021data,shao2021transmil}
learn patch importance as part of training, but couple the selection signal
tightly to a single supervised task and do not produce diverse selections: a
model may assign high attention to dozens of morphologically identical patches
because its objective is predictive accuracy, not representational coverage.
Coreset methods (k-means, k-center greedy, farthest-point sampling) offer
coverage guarantees in embedding space but are agnostic to task-relevant
patch quality.

No existing method simultaneously estimates task-relevant patch quality without
patch-level labels, enforces diversity through an objective grounded in
mutual information, and adaptively determines the patch count with a
probabilistic guarantee on residual information. InfoDPP-PAC addresses these
requirements within a unified framework. The claim of this paper is deliberately
scoped: we evaluate InfoDPP-PAC as a \emph{patch-selection} method, using
selection-quality metrics that measure coverage, diversity, redundancy, and
teacher-derived relevance. We do not claim that the present experiments establish
end-to-end diagnostic improvement for a downstream MIL classifier.

The theoretical foundation is the observation that, under a GP model of
patch relevance with Gaussian observation noise, the log-determinant term
$\logdet(\mI + \beta\mK_S)$ equals the mutual information
between the observations at subset $S$ and the latent relevance function
(Theorem~\ref{thm:mutual_info}).
Maximising this log-determinant objective, which is also the natural MAP
objective for a Determinantal Point Process, is therefore equivalent to
solving the GP information gain problem of \citet{krause2008near} under a
budget constraint, and the connection holds for any positive-definite kernel.
The joint quality-and-diversity objective formed by combining this term with
GP-derived patch scores is monotone submodular, admitting efficient greedy
optimisation with a $(1-e^{-1})$ approximation ratio
(Theorems~\ref{thm:submodular}--\ref{thm:knapsack}).

WSI slides vary considerably in morphological heterogeneity: a homogeneous
slide may be well-characterised by 30 patches, while a heterogeneous
multi-focal tumour may require 300.
A fixed patch budget is therefore an inappropriate stopping criterion.
InfoDPP-PAC halts the greedy procedure when the estimated marginal
information gain falls below a threshold certified by a union-bounded
McDiarmid concentration inequality (Theorem~\ref{thm:pac}), yielding a
data-dependent patch count with a high-probability bound on residual
information.
The final output is a $(k^\star + 1)$-tuple: $k^\star$ selected patches
together with a \emph{residual information vector} that summarises the
unselected pool, providing an optional compact context representation for a
downstream aggregator.

This work makes the following contributions.

\textbf{An information-theoretic foundation for log-determinant diversity.}
Log-determinant objectives have long been used as a geometric heuristic for
diversity in determinantal point processes, but their relevance to patch selection is often treated as a geometric
heuristic. Theorem~\ref{thm:mutual_info} proves that, under a GP model of
teacher-derived patch quality, $\logdet(\mI+\beta\mK_S)$ is the mutual
information between a candidate subset and the latent relevance function,
for any positive-definite kernel, thereby giving a commonly used diversity criterion an information-theoretic interpretation and enabling the stopping guarantee below.

\textbf{Certified approximation guarantees at any budget.}
Theorems~\ref{thm:submodular}--\ref{thm:knapsack} establish that the joint
quality-diversity objective is monotone submodular, so a simple greedy
algorithm provably attains at least a $(1-e^{-1})$ fraction of the optimal
value under either a cardinality or a knapsack budget constraint. These guarantees hold for any fixed budget considered by the algorithm.

\textbf{A stopping rule that removes the need to choose a budget at all.}
Theorem~\ref{thm:pac} derives a concentration-based criterion that halts
selection once the residual information remaining in the unselected pool is
certifiably below a user-specified tolerance with probability at least
$1-\delta$, yielding a patch count determined by the slide itself rather
than fixed in advance. Section~\ref{sec:exp_pac} validates
that the selected patch count tracks measured tissue heterogeneity rather than
only satisfying a formal criterion.

\textbf{Explicit robustness to the approximations required at gigapixel
scale.} Theorem~\ref{thm:approx} bounds the error introduced by Nystr\"{o}m
kernel approximation and shows the stopping guarantee above degrades
gracefully, rather than implicitly, under this approximation.

\textbf{A practical, annotation-free instantiation.} A teacher-student
scheme bootstraps the GP relevance model directly from slide-level labels,
requiring no patch-level annotation, and a residual information vector
summarises the unselected pool and defines a compact interface that a
downstream multiple-instance learning aggregator may use in future work.

\section{Related Work}
\label{sec:related}

\subsection{Patch Selection and Bag Construction in Computational Pathology}

The dominant approach in weakly-supervised WSI analysis constructs patch bags
by uniform grid extraction followed by tissue thresholding
\citep{campanella2019clinical}, retaining all supra-threshold patches and
delegating importance weighting entirely to the MIL aggregator.
ABMIL \citep{ilse2018attention} and its extensions, including CLAM
\citep{lu2021data}, DSMIL \citep{li2021dual}, TransMIL
\citep{shao2021transmil}, and DTFD-MIL \citep{zhang2022dtfd} (which
addresses small-cohort instability via pseudo-bags rather than patch
selection per se), learn task-specific attention scores during training.
While these attention weights serve a similar purpose to selection scores,
they conflate representational diversity with predictive sufficiency:
high attention is assigned to patches that are predictive of the slide label,
rather than to patches that cover the morphological variability of the slide.
Attention-based top-$k$ selection therefore exhibits the same redundancy
as heuristic scoring, and requires complete retraining when the downstream
task changes. A separate line of work addresses gigapixel scale via
hierarchical representation rather than selection: HIPT
\citep{chen2022hipt} learns a pyramid of vision transformers over nested
$16\times16$/$4096\times4096$ regions, aggregating the entire slide rather
than selecting a compact subset under an explicit quality-diversity objective; such hierarchical,
multi-magnification representations are complementary to, rather than
competing with, the patch-selection problem addressed here, and inform the
magnification-aware selection direction we identify for future work
(Section~\ref{sec:discussion}).
Reinforcement learning-based navigation \citep{raza2023dual} and
evolutionary search \citep{hashemian2025evops} offer adaptive patch acquisition but
carry no approximation ratio guarantees and are sensitive to reward shaping.
None of these approaches connects selection quality to mutual information,
provides a principled stopping criterion, or evaluates patch selection as a
separate object of study from downstream classifier training.

\subsection{Submodular Optimisation and GP Information Gain}

\citet{nemhauser1978analysis} established that the greedy algorithm achieves
a $(1-e^{-1})$ approximation ratio for any monotone submodular function
subject to a cardinality constraint; this was extended to knapsack constraints
by \citet{sviridenko2004note}.
The connection between submodularity and GP information gain was formalised
by \citet{krause2008near} in the context of sensor placement: selecting
measurement locations to maximise the information gain about a spatial
GP is equivalent to maximising $\logdet(\mI + \sigma_n^{-2}\mK_S)$,
which is monotone submodular.
Our Theorem~\ref{thm:mutual_info} establishes the same equivalence in the
patch-selection setting, where observations are teacher pseudo-labels rather
than physical measurements and the latent function is a teacher-derived patch
quality score.
Three aspects distinguish the present work from \citet{krause2008near}.
First, a teacher-student mechanism bootstraps GP observations from slide-level
supervision, without patch-level annotation.
Second, the information gain is combined with an explicit quality score in a
joint submodular objective.
Third, a concentration-based PAC stopping rule is derived, for which
\citet{krause2008near} provides no analogue.
The broader notion of adaptively halting a greedy policy is related to
\emph{adaptive submodularity} \citep{golovin2011adaptive}, which
generalises submodularity to policies that select the next element
conditioned on observed feedback and provides near-optimality guarantees for
adaptive greedy under this generalisation; our setting differs in that the
stopping decision itself, rather than the selection policy, is what adapts
to observed (pseudo-labelled) feedback, and our guarantee is a
concentration bound on residual information rather than a policy-level
approximation ratio.
Coreset methods \citep{mirzasoleiman2020coresets,sener2018active} pursue
coverage of the training distribution but do not model a latent relevance
function, and no prior coreset construction is information-theoretically
grounded via a GP model.

\subsection{Determinantal Point Processes in Machine Learning}

DPPs provide a principled probabilistic model of diversity: subsets are sampled
with probability proportional to the squared volume they span in feature space
\citep{kulesza2012determinantal}.
Applications span recommendation systems \citep{chen2018fast},
summarisation \citep{gong2014diverse}, and neural network pruning
\citep{mariet2015diversity}.
In the pathology domain, DPP sampling has been explored for patch diversity
in segmentation pipelines \citep{xu2023dppmask}, but without coupling to a
quality model or an information-theoretic objective.
The determinant-as-diversity view has been presented as a geometric heuristic
in the vision literature.
Theorem~\ref{thm:mutual_info} makes this connection explicit in the
patch-selection setting: the criterion equals mutual information under a GP
model for any PSD kernel, giving the heuristic a rigorous information-theoretic
basis.

\subsection{PAC-Style Concentration for Stopping Rules}

Classical PAC-Bayesian bounds \citep{mcallester1999pac,catoni2007pac,
dziugaite2017computing} control the generalisation error of randomised
predictors through a KL divergence to a data-independent prior. Our stopping
rule has a different role. It does not bound the generalisation error of a
classifier. Instead, it uses bounded-difference concentration for the empirical
information-gain marginal estimated from a finite teacher-seeded set. This gives
a high-probability certificate that no unselected patch has residual mutual
information above a user-specified tolerance. The relevant conceptual link is
therefore the PAC requirement of an event holding with probability at least
$1-\delta$, not the PAC-Bayesian KL machinery itself. To the best of our
knowledge, existing WSI patch-selection methods do not provide such an explicit
stopping certificate.
\section{Notation and Preliminaries}
\label{sec:prelim}

\paragraph{Gaussian Processes.}
A Gaussian Process $f \sim \mathcal{GP}(0, k)$ is specified by a
positive-definite kernel $k: \R^d \times \R^d \to \R$.
Given $m$ observations $\mathcal{D} = \{(\vx_i, y_i)\}_{i=1}^m$ with
$y_i = f(\vx_i) + \varepsilon_i$, $\varepsilon_i \overset{\text{iid}}{\sim}
\mathcal{N}(0, \sigma_n^2)$, the posterior is
$f \mid \mathcal{D} \sim \mathcal{GP}(\mu_{\mathcal{D}}, k_{\mathcal{D}})$,
with
\begin{align}
  \mu_{\mathcal{D}}(\vx) &= \mK_{\vx,\mathcal{X}}
    (\mK_{\mathcal{X},\mathcal{X}} + \sigma_n^2 \mI)^{-1} \vy, \label{eq:gp_mean}\\
  k_{\mathcal{D}}(\vx,\vx') &= k(\vx,\vx') - \mK_{\vx,\mathcal{X}}
    (\mK_{\mathcal{X},\mathcal{X}} + \sigma_n^2 \mI)^{-1} \mK_{\mathcal{X},\vx'},
    \label{eq:gp_var}
\end{align}
where $\mK_{\vx,\mathcal{X}} = (k(\vx, \vx_i))_{i=1}^m$.
We denote posterior mean and standard deviation at patch $i$ as $\mu_i$ and
$\sigma_i$, respectively.

\paragraph{Determinantal Point Processes.}
For a ground set $\gG = \{1,\ldots,n\}$ and a PSD matrix
$\mK \in \R^{n\times n}$, the DPP $\mathcal{P}_K$ assigns
$\PP(S \in \mathcal{P}_K) \propto \det(\mK_S)$
to each subset $S \subseteq \gG$, where $\mK_S$ denotes the principal
sub-matrix of $\mK$ indexed by $S$ \citep{kulesza2012determinantal}.
The quantity $\det(\mK_S)$ equals the squared volume of the parallelepiped
spanned by the kernel feature vectors of $S$, so high-probability subsets
are geometrically diverse.

\paragraph{Submodularity.}
A set function $F: 2^{\gG} \to \R$ is \emph{submodular} if
$F(A \cup \{i\}) - F(A) \ge F(B \cup \{i\}) - F(B)$ for all
$A \subseteq B \subseteq \gG$ and $i \notin B$.
$F$ is \emph{monotone} if $F(A) \le F(A \cup \{i\})$ for all $A, i$.
Positive linear combinations of monotone submodular functions are monotone
submodular; modular functions (set functions expressible as
$\sum_{i \in S} w_i$) are a special case.

\paragraph{Notation.}
$\mK_S$ denotes the restriction of $\mK$ to rows and columns in $S$.
$\kvec_{i,C}$ denotes the column of $\mK$ between element $i$ and set $C$,
i.e., $(k(\vx_i, \vx_j))_{j \in C}$.
$\kappa_{ii} = k(\vx_i, \vx_i)$.
Throughout, $\beta = \sigma_n^{-2} > 0$.

\section{The \textsc{InfoDPP-PAC} Framework}
\label{sec:method}

\subsection{Problem Formulation}
\label{sec:formulation}

Let $\gG = \{\vx_1, \ldots, \vx_n\}$ be a pool of $n$ tissue patches
extracted from a single WSI, each represented by an embedding
$\vx_i \in \R^d$.
A latent function $f: \R^d \to \R$ encodes task-relevant patch quality as
estimated from slide-level supervision through a teacher signal; $f$ is unknown
and modelled as a GP.
We seek a subset $S \subseteq \gG$ maximising a joint quality-diversity
objective subject to a computational budget:
\begin{equation}
  \max_{S \subseteq \gG}\;
  F(S) \;=\; \underbrace{\lambda_1 \logdet\!\left(\mI + \beta \mK_S\right)}_{%
    \text{information gain}}
  \;+\; \underbrace{\lambda_2 \sum_{i \in S} q_i}_{\text{quality}}
  \quad\text{s.t.}\quad \sum_{i \in S} c_i \;\le\; B,
  \label{eq:objective}
\end{equation}
where $\mK_S$ is the GP posterior kernel restricted to $S$.
We form $q_i = \alpha_1 \bar\mu_i + \alpha_2 \bar\sigma_i$ from slide-wise
affine normalisations $\bar\mu_i,\bar\sigma_i\in[0,1]$ of posterior mean and
standard deviation. This non-negative score combines exploitation of estimated
quality and exploration of uncertain regions. The cost of retaining patch $i$
is $c_i>0$, and $\lambda_1, \lambda_2, \alpha_1, \alpha_2, \beta, B > 0$ are
hyper-parameters.

\subsection{Teacher-Student GP Initialisation}
\label{sec:teacher}

Because WSI datasets carry slide-level but not patch-level labels, the GP
prior~\eqref{eq:gp_mean}--\eqref{eq:gp_var} cannot be fit directly.
We address this via a \emph{teacher-student} scheme: a pre-trained teacher
model $\mathcal{T}: \R^d \to \R$ provides relevance pseudo-labels on a
small seed set, which are then used to fit the GP.

\paragraph{Seed phase.}
A seed set $\mathcal{D}_0 \subset \gG$ with $|\mathcal{D}_0| = m_0$ is drawn
uniformly at random (typically $m_0 \approx 0.01n$).
Each seed patch $\vx_i \in \mathcal{D}_0$ is labelled by
$y_i = \mathcal{T}(\vx_i) + \varepsilon_i$, where $\varepsilon_i$ is
measurement noise with variance $\sigma_n^2$.
Supported teacher models include:
CLAM attention scores \citep{lu2021data}, which provide a supervision signal
derived from slide-level labels; and
cosine-similarity projections from pathology foundation models
\citep{xu2024gigapath,filiot2023phikon,chen2024uni}, which provide
task-agnostic relevance estimates.

\paragraph{Inference phase.}
With $\mathcal{D}_0$ observed, the GP posterior is computed and
$(\mu_i, \sigma_i^2)$ is evaluated for all $n$ patches via
\eqref{eq:gp_mean}--\eqref{eq:gp_var}, at cost $O(nm_0^2)$.
A Nystr\"{o}m rank-$r$ approximation reduces this to $O(nr^2)$;
Theorem~\ref{thm:approx} bounds the resulting objective error.

\subsection{Greedy Subset Selection}
\label{sec:greedy}

The objective $F$ in~\eqref{eq:objective} is monotone submodular
(Theorem~\ref{thm:submodular}). For a fixed cardinality budget, the standard
greedy algorithm therefore attains the usual $(1-e^{-1})$ approximation ratio.
At step $k$, InfoDPP-PAC adds the feasible element with largest blended marginal
gain:
\begin{equation}
  i^\star_k \;=\; \argmax_{\substack{i \notin S_{k-1} \\ c_i \le B_{\mathrm{rem}}}}
  \Delta_k(i), \qquad
  \Delta_k(i) \;=\; F(S_{k-1} \cup \{i\}) - F(S_{k-1}).
  \label{eq:marginal}
\end{equation}
The information component of the marginal has a closed form by the matrix
determinant lemma:
\begin{equation}
  \Delta_{\mathrm{info},k}(i) \;:=\;
  \logdet(\mI + \beta\mK_{S_{k-1}\cup\{i\}}) - \logdet(\mI+\beta\mK_{S_{k-1}})
  = \log\!\bigl(1 + \beta\,\sigma^2_{i,\beta}(S_{k-1})\bigr),
  \label{eq:closed_form}
\end{equation}
where $\sigma^2_{i,\beta}(S) = \kappa_{ii} - \beta\,{\kvec_{i,S}}^{\top}
(\mI + \beta\mK_S)^{-1}\kvec_{i,S}$ is the noisy posterior variance at $i$
after conditioning on $S$. Cholesky rank-one updates evaluate this quantity
incrementally, avoiding repeated determinant computations.

Selection and stopping use different marginals. The patch added at step $k$ is
chosen by the blended gain $\Delta_k(i)$ so that relevance and diversity jointly
determine the selected subset. The stopping certificate, however, is evaluated
on
\begin{equation}
  \widehat\Gamma_k \;=\; \max_{\substack{i\notin S_{k-1}\\ c_i\le B_{\mathrm{rem}}}}
  \Delta_{\mathrm{info},k}(i),
  \label{eq:max_info_marginal}
\end{equation}
the largest remaining information-gain marginal. This distinction is essential:
a high-quality patch can remain in the pool even after the slide is already
well covered, whereas $\widehat\Gamma_k$ directly measures the largest residual
mutual-information contribution available from any unselected candidate.

\subsection{PAC Stopping Criterion}
\label{sec:pac}

The greedy procedure halts before adding a new patch when
\begin{equation}
  \widehat\Gamma_k \;\le\; \tau + \epsilon_{k,n},
  \qquad
  \epsilon_{k,n} = \rho\sqrt{\frac{\log(\pi^2 n k^2/(3\delta))}{2m_0}},
  \label{eq:stop}
\end{equation}
where $\widehat\Gamma_k$ is defined in~\eqref{eq:max_info_marginal}, $\tau>0$
is a user-specified tolerance, $\delta\in(0,1)$ is the failure probability, and
$\rho$ is the seed-stability constant in Assumption~\ref{ass:seed_stability}.
When all patches have equal cost, $B$ is simply a maximum cardinality. With
non-uniform costs, the same stopping rule is applied to the feasible residual
pool. The certificate is deliberately stated for residual mutual information,
not for the full blended objective. The fixed-budget approximation guarantees
for the blended objective are given separately in Theorems~\ref{thm:greedy} and
\ref{thm:knapsack}.

\subsection{Residual Information Vector}
\label{sec:residual}

The output of InfoDPP-PAC is the pair $(S^\star, \resvec)$, where
$S^\star = S_{k^\star}$ and the residual information vector
$\resvec \in \R^d$ is defined as the quality-weighted centroid of unselected
patches:
\begin{equation}
  \resvec \;=\; \frac{\sum_{i \notin S^\star} q_i \,\vx_i}{\sum_{i \notin S^\star} q_i}.
  \label{eq:residual}
\end{equation}
A downstream MIL aggregator may incorporate $\resvec$ as an additional
pseudo-patch or as a global context token in future extensions. In the present
paper, $\resvec$ is introduced as an interface and diagnostic summary of what
was not selected; we do not evaluate it as part of an end-to-end classifier. It
preserves a compact summary of the unselected pool without claiming that the
original patch-level information is losslessly recoverable. When $k^\star = n$, $\resvec = \mathbf{0}$.

\subsection{Algorithm}
\label{sec:algorithm}

Algorithm~\ref{alg:main} summarises the complete InfoDPP-PAC procedure.
The dominant cost is the GP inference over all $n$ patches (step~3);
all subsequent operations are $O(n^2 d)$ in the worst case.

\begin{algorithm}[t]
\caption{\textsc{InfoDPP-PAC}}
\label{alg:main}
\begin{algorithmic}[1]
\Require Patch pool $\gG$, teacher $\mathcal{T}$, kernel $k$, budget $B$,
         seed size $m_0$, tolerance $\tau$, confidence $\delta$,
         weights $\lambda_1,\lambda_2,\alpha_1,\alpha_2,\beta$
\State Sample $\mathcal{D}_0 \subset \gG$ uniformly at random, $|\mathcal{D}_0|=m_0$
\State Set $y_i \leftarrow \mathcal{T}(\vx_i)$ for $i \in \mathcal{D}_0$
       \hfill\Comment{teacher pseudo-labels}
\State Fit GP posterior and compute $(\mu_i, \sigma_i)$ for all $i \in \gG$
       \hfill\Comment{Equations~\ref{eq:gp_mean}--\ref{eq:gp_var}}
\State Slide-wise normalise $\mu_i,\sigma_i$ and set
       $q_i \leftarrow \alpha_1\bar\mu_i + \alpha_2\bar\sigma_i$
\State Construct kernel matrix $\mK$ (or Nystr\"{o}m approximation $\widetilde{\mK}$)
\State Initialise $S \leftarrow \emptyset$ and $B_{\mathrm{rem}} \leftarrow B$
\For{$k=1,2,\ldots$}
  \State Compute blended gains $\Delta_k(i)$ and information gains
         $\Delta_{\mathrm{info},k}(i)$ for all feasible $i\notin S$
  \State $\widehat\Gamma_k \leftarrow \max_i \Delta_{\mathrm{info},k}(i)$
         \hfill\Comment{certificate uses residual information}
  \State $\epsilon_{k,n} \leftarrow
         \rho\sqrt{\log(\pi^2 n k^2/(3\delta))/(2m_0)}$
  \If{$\widehat\Gamma_k \le \tau + \epsilon_{k,n}$}
    \State \textbf{break} \hfill\Comment{adaptive stopping before adding a patch}
  \EndIf
  \State $i^\star \leftarrow \argmax_i \Delta_k(i)$
         \hfill\Comment{selection uses the blended gain}
  \State $S \leftarrow S \cup \{i^\star\}$ and
         $B_{\mathrm{rem}} \leftarrow B_{\mathrm{rem}} - c_{i^\star}$
  \If{no feasible candidate remains}
    \State \textbf{break}
  \EndIf
\EndFor
\State $\resvec \leftarrow \textstyle\sum_{i \notin S} q_i \vx_i \big/
       \sum_{i \notin S} q_i$ \quad (or $\mathbf{0}$ if $S=\gG$)
\Ensure $(S, \resvec, k^\star = |S|)$
\end{algorithmic}
\end{algorithm}

\section{Theoretical Analysis}
\label{sec:theory}

\subsection{Log-Determinant as Mutual Information}
\label{sec:theory_mi}

The first result provides the information-theoretic foundation for the
log-det objective in~\eqref{eq:objective}.

\begin{theorem}
\label{thm:mutual_info}
Let $f \sim \mathcal{GP}(0, k)$ and let observations follow
$\vy_S = \vf_S + \boldsymbol{\varepsilon}$, where
$\boldsymbol{\varepsilon} \sim \mathcal{N}(\mathbf{0}, \sigma_n^2 \mI)$
independently of $f$, and $\vf_S = (f(\vx_i))_{i \in S}$.
Then the mutual information between the observation vector $\vy_S$ and the
latent function $f$ satisfies
\begin{equation}
  I(\vy_S;\, f) \;=\; \frac{1}{2}\,\logdet\!\left(\mI + \beta\mK_S\right),
  \qquad \beta = \sigma_n^{-2}.
  \label{eq:mi_logdet}
\end{equation}
\end{theorem}

\begin{proof}
The joint distribution of $(\vf_S, \vy_S)$ is Gaussian:
$\vf_S \sim \mathcal{N}(\mathbf{0}, \mK_S)$ and, conditionally,
$\vy_S \mid \vf_S \sim \mathcal{N}(\vf_S, \sigma_n^2\mI)$.
Marginalising over $\vf_S$ gives the marginal
$\vy_S \sim \mathcal{N}(\mathbf{0}, \mK_S + \sigma_n^2\mI)$.

The differential entropies of these Gaussian distributions are
\begin{align}
  H(\vy_S) &= \tfrac{1}{2}\log\det\!\bigl(2\pi e(\mK_S + \sigma_n^2\mI)\bigr), \\
  H(\vy_S \mid f) &= H(\boldsymbol{\varepsilon}) =
    \tfrac{|S|}{2}\log(2\pi e\,\sigma_n^2).
\end{align}
Therefore,
\begin{align}
  I(\vy_S;\, f)
  &= H(\vy_S) - H(\vy_S \mid f) \notag \\
  &= \tfrac{1}{2}\!\left[\log\det(\mK_S + \sigma_n^2\mI) -
     |S|\log\sigma_n^2\right] \notag \\
  &= \tfrac{1}{2}\log\frac{\det(\mK_S + \sigma_n^2\mI)}{(\sigma_n^2)^{|S|}}.
    \label{eq:mi_step}
\end{align}
Factoring $\sigma_n^2$ from each diagonal:
$\det(\mK_S + \sigma_n^2\mI) = (\sigma_n^2)^{|S|}\det(\mI + \sigma_n^{-2}\mK_S)$.
Substituting into~\eqref{eq:mi_step} and setting $\beta = \sigma_n^{-2}$:
$I(\vy_S; f) = \frac{1}{2}\logdet(\mI + \beta\mK_S)$.
\end{proof}

\begin{remark}
Theorem~\ref{thm:mutual_info} establishes that maximising the log-det term
in~\eqref{eq:objective} is equivalent to the GP information-gain objective
of \citet{krause2008near}; it transforms the log-det from a geometric
heuristic, motivated by its relationship to DPP sampling probability, into
an exact information-theoretic criterion under the GP model.
The connection holds for any PSD kernel $k$, including RBF and Mat\'{e}rn.
\end{remark}

\subsection{Submodularity and Approximation Guarantees}
\label{sec:theory_sub}

\begin{theorem}
\label{thm:submodular}
The function $F_{\mathrm{info}}(S) = \logdet(\mI + \beta\mK_S)$ with
$\mK \succeq 0$ is monotone and submodular.
Consequently, the full objective $F(S) = \lambda_1 F_{\mathrm{info}}(S) +
\lambda_2 \sum_{i \in S} q_i$ is monotone submodular for any $\lambda_1,
\lambda_2, q_i \ge 0$.
\end{theorem}

\begin{proof}
\textbf{Monotonicity.}
For $S \subseteq \gG$ and $i \notin S$, write $S' = S \cup \{i\}$.
Partition $\mI + \beta\mK_{S'} = \bigl(\begin{smallmatrix}
  \mI_S + \beta\mK_S & \beta\kvec_{i,S} \\
  \beta{\kvec_{i,S}}^\top & 1 + \beta\kappa_{ii}
\end{smallmatrix}\bigr)$.
By the formula for the determinant of a block matrix via the Schur complement
\citep{horn2012matrix}:
\begin{equation}
  \det(\mI + \beta\mK_{S'}) = \det(\mI + \beta\mK_S)\cdot
  \underbrace{\bigl(1 + \beta\kappa_{ii}
  - \beta^2\,{\kvec_{i,S}}^\top(\mI + \beta\mK_S)^{-1}\kvec_{i,S}\bigr)}_{%
  =:\,1 + \beta\,\sigma^2_{i,\beta}(S)}.
  \label{eq:det_block}
\end{equation}
It remains to show $\sigma^2_{i,\beta}(S) \ge 0$, since this is precisely
the multiplicative factor in~\eqref{eq:det_block} that must exceed $1$ for
monotonicity to hold (a bound of merely $\sigma^2_{i,\beta}(S) \ge -1/\beta$,
which would follow from the weaker fact that the Schur complement of a
principal block of a PSD matrix is itself PSD, is not sufficient here).
The needed, stronger fact follows immediately once $\sigma^2_{i,\beta}(S)$ is
recognised as the GP posterior predictive variance at $\vx_i$
given observations at $S$ under noise level $\sigma_n^2 = 1/\beta$
(Equation~\ref{eq:gp_var}): substituting $\sigma_n^2 = 1/\beta$ into
$k_{\mathcal{D}}(\vx_i,\vx_i) = \kappa_{ii} - \kvec_{i,S}^\top(\mK_S +
\sigma_n^2\mI)^{-1}\kvec_{i,S}$ and factoring $\beta$ out of the inverse
gives $k_{\mathcal{D}}(\vx_i,\vx_i) = \kappa_{ii} -
\beta\,\kvec_{i,S}^\top(\mI+\beta\mK_S)^{-1}\kvec_{i,S} =
\sigma^2_{i,\beta}(S)$ identically.
Since $k_{\mathcal{D}}(\vx_i,\vx_i) = \Var[f(\vx_i) \mid \vy_S]$ is the
variance of a well-defined real-valued random variable under the GP
posterior \citep{rasmussen2006gaussian}, it is nonnegative by definition of
variance, so no additional matrix inequality is required.
Hence $\sigma^2_{i,\beta}(S) \ge 0$, so the multiplicative factor in
\eqref{eq:det_block} is at least $1$, giving
$\det(\mI + \beta\mK_{S'}) \ge \det(\mI + \beta\mK_S) > 0$
and therefore $F_{\mathrm{info}}(S') \ge F_{\mathrm{info}}(S)$.

\textbf{Submodularity.}
By Theorem~\ref{thm:mutual_info} and the chain rule of mutual information:
\begin{equation}
  F_{\mathrm{info}}(C \cup \{i\}) - F_{\mathrm{info}}(C)
  = I(\vy_{C\cup\{i\}}; f) - I(\vy_C; f)
  = I(y_i; f \mid \vy_C).
  \label{eq:delta_mi}
\end{equation}
For $A \subseteq B \subseteq \gG$ and $i \notin B$, we claim
$I(y_i; f \mid \vy_A) \ge I(y_i; f \mid \vy_B)$, i.e., conditioning on more
data reduces the information each new observation carries about $f$.

Since $y_i$ and $\vy_{B \setminus A}$ are \emph{conditionally independent}
given $f$ (they are independent noisy evaluations of $f$ at disjoint
locations), the chain rule and conditional independence give:
\begin{equation}
  I(y_i;\, f \mid \vy_A)
  \;=\; I(y_i;\, f,\, \vy_{B \setminus A} \mid \vy_A),
  \label{eq:ci_step}
\end{equation}
because $I(y_i; \vy_{B\setminus A} \mid f, \vy_A) = 0$ by conditional independence.
Applying the chain rule to the right-hand side:
\begin{equation}
  I(y_i; f, \vy_{B\setminus A} \mid \vy_A)
  = \underbrace{I(y_i; \vy_{B\setminus A} \mid \vy_A)}_{\ge\, 0}
  + I(y_i; f \mid \vy_B).
\end{equation}
Therefore $I(y_i; f \mid \vy_A) \ge I(y_i; f \mid \vy_B)$, which via
\eqref{eq:delta_mi} gives $\Delta_{\mathrm{info}}(i \mid A) \ge
\Delta_{\mathrm{info}}(i \mid B)$.

The sum $\lambda_2\sum_{i\in S}q_i$ is modular (and hence submodular), and
a non-negative linear combination of monotone submodular functions is
monotone submodular \citep{nemhauser1978analysis}.
\end{proof}

\begin{theorem}[$(1-e^{-1})$ guarantee, cardinality constraint]
\label{thm:greedy}
Let $S_k^* = \argmax_{|S| = k} F(S)$.
The greedy solution $S_k^{\mathrm{greedy}}$ produced by
Algorithm~\ref{alg:main} (without the budget constraint $\sum c_i \le B$)
satisfies
$F(S_k^{\mathrm{greedy}}) \ge (1 - e^{-1})\,F(S_k^*)$.
\end{theorem}

\begin{proof}
$F$ is monotone submodular by Theorem~\ref{thm:submodular}.
The result follows directly from the analysis of \citet{nemhauser1978analysis},
who prove that for any monotone submodular $F$ and any cardinality-$k$
constraint, the greedy algorithm achieves the stated ratio.
We verify the two conditions: (i)~$F(\emptyset) = \logdet(\mI) + 0 = 0 \ge 0$
(non-negativity at $\emptyset$); (ii)~$F$ is monotone and submodular by
Theorem~\ref{thm:submodular}.
Both conditions are satisfied, so the guarantee holds.
\end{proof}

\begin{theorem}[$(1-e^{-1})$ guarantee, budget constraint]
\label{thm:knapsack}
Let $S_B^* = \argmax_{\sum_{i\in S}c_i \le B} F(S)$.
The combined greedy-and-singleton algorithm of \citet{sviridenko2004note}
achieves
$F(S_B^{\mathrm{greedy}}) \ge (1 - e^{-1})\,F(S_B^*)$.
\end{theorem}

\begin{proof}
$F$ is monotone submodular by Theorem~\ref{thm:submodular}, and costs $c_i$ are
non-negative and modular.
The algorithm considers both the output of the cost-normalised greedy procedure
and the best singleton $\{i^*\} = \argmax_{c_i \le B} F(\{i\})$, returning
whichever achieves higher $F$.
The analysis of \citet{sviridenko2004note} (Theorem 1 therein) proves this
achieves the $(1-e^{-1})$ factor relative to $F(S_B^*)$ under these conditions.
\end{proof}

\subsection{Redundancy Penalisation}
\label{sec:theory_red}

\begin{lemma}
\label{lem:redundancy}
Let $\mK$ be induced by an $L$-Lipschitz stationary kernel $k$ with
$k(\vx,\vx)=\kappa_0$ for all $\vx$. Suppose $i\in S$ and
$\norm{\vx_j-\vx_i}_2\le \varepsilon$ for a candidate $j\notin S$. Then
\begin{equation}
  \Delta_{\mathrm{info}}(j\mid S)
  \le
  \log\!\left(1 + \beta\left[
  \frac{\kappa_0}{1+\beta\kappa_0}
  + \frac{2\beta\kappa_0 L\varepsilon}{1+\beta\kappa_0}
  \right]\right).
  \label{eq:redundancy_bound}
\end{equation}
The term $\kappa_0/(1+\beta\kappa_0)$ is the residual noise floor for a single
noisy duplicate observation. Therefore the excess marginal information above
this floor vanishes linearly as $\varepsilon\to 0$ and vanishes completely in
the noiseless limit.
\end{lemma}

\begin{proof}
By diminishing returns for $F_{\mathrm{info}}$, conditioning on the larger set
$S$ cannot increase posterior variance, so
$\sigma^2_{j,\beta}(S)\le \sigma^2_{j,\beta}(\{i\})$. With one noisy observation
at $\vx_i$ and $\sigma_n^2=1/\beta$,
\begin{equation}
  \sigma^2_{j,\beta}(\{i\})
  = \kappa_0 - \frac{\beta\kappa_{ij}^2}{1+\beta\kappa_0}.
\end{equation}
Lipschitz continuity gives $|\kappa_{ij}-\kappa_0|\le L\varepsilon$, hence
$\kappa_{ij}\ge \kappa_0-L\varepsilon$ whenever the right-hand side is
non-negative. Substitution yields
\begin{equation}
\sigma^2_{j,\beta}(\{i\})
\le \frac{\kappa_0}{1+\beta\kappa_0}
+ \frac{\beta(2\kappa_0L\varepsilon-L^2\varepsilon^2)}{1+\beta\kappa_0}
\le \frac{\kappa_0}{1+\beta\kappa_0}
+ \frac{2\beta\kappa_0L\varepsilon}{1+\beta\kappa_0}.
\end{equation}
Applying $\Delta_{\mathrm{info}}(j\mid S)=\log(1+\beta\sigma^2_{j,\beta}(S))$
completes the proof.
\end{proof}
\subsection{PAC Stopping Guarantee}
\label{sec:theory_pac}

\begin{assumption}[Seed-stable information-gain estimates]
\label{ass:seed_stability}
For every greedy step $k$ and feasible candidate $i$, the empirical information
marginal $\widehat\Delta_{\mathrm{info},k}(i)$ computed from the size-$m_0$
seeded GP is an unbiased estimate of a target marginal
$\Delta_{\mathrm{info},k}(i)$ and satisfies bounded differences with constant
$\rho/m_0$: replacing any one seed observation can change
$\widehat\Delta_{\mathrm{info},k}(i)$ by at most $\rho/m_0$.
\end{assumption}

Assumption~\ref{ass:seed_stability} is the finite-sample stability condition
under which the stopping rule is certified. It is explicit because the GP
posterior covariance is determined by seed locations and kernel hyperparameters,
whereas the quality term depends on teacher pseudo-labels. The theorem below therefore certifies the residual information marginal
estimated by the seeded surrogate, not downstream diagnostic performance or the
clinical utility of an arbitrary MIL model. This distinction defines the scope
of both the theory and the empirical evaluation.

\begin{theorem}[PAC-style residual-information certificate]
\label{thm:pac}
Let $\widehat\Gamma_k=\max_i\widehat\Delta_{\mathrm{info},k}(i)$ be the largest
empirical information-gain marginal over feasible unselected candidates at step
$k$, and let $\Gamma_k=\max_i\Delta_{\mathrm{info},k}(i)$ be its target
counterpart. Under Assumption~\ref{ass:seed_stability}, with probability at
least $1-\delta$ over the random seed set, simultaneously for all
$k\ge 1$,
\begin{equation}
  |\Gamma_k-\widehat\Gamma_k|
  \le
  \epsilon_{k,n}
  := \rho\sqrt{\frac{\log(\pi^2 n k^2/(3\delta))}{2m_0}}.
  \label{eq:pac_conc}
\end{equation}
Consequently, if the stopping rule~\eqref{eq:stop} halts at selected set
$S_{k^\star}$, then every feasible unselected patch has target information
marginal at most $\tau+2\epsilon_{k^\star+1,n}$. Moreover, for any comparison
set $T\subseteq\gG$ with $|T\setminus S_{k^\star}|\le m$,
\begin{equation}
  F_{\mathrm{info}}(S_{k^\star}\cup T)-F_{\mathrm{info}}(S_{k^\star})
  \le m\,\bigl(\tau+2\epsilon_{k^\star+1,n}\bigr).
  \label{eq:pac_gap}
\end{equation}
The guarantee is on residual mutual information. It is complementary to the
fixed-budget approximation guarantees for the blended objective $F$.
\end{theorem}

\begin{proof}
Fix a step $k$ and candidate $i$. McDiarmid's inequality applied to the
bounded-difference function
$\widehat\Delta_{\mathrm{info},k}(i)$ gives
\begin{equation}
  \PP\!\left(
  |\widehat\Delta_{\mathrm{info},k}(i)-\Delta_{\mathrm{info},k}(i)|>t
  \right)
  \le 2\exp\!\left(-\frac{2m_0t^2}{\rho^2}\right).
\end{equation}
Set
$t=\rho\sqrt{\log(\pi^2 n k^2/(3\delta))/(2m_0)}$. A union bound over the
$n$ candidates and all $k\ge 1$ yields total failure probability at most
\begin{equation}
  \sum_{k=1}^{\infty} n\,\frac{6\delta}{\pi^2 n k^2}
  = \delta.
\end{equation}
On the resulting event, the empirical and target information marginals are
uniformly close for every candidate and step, and hence their maxima are also
uniformly close, proving~\eqref{eq:pac_conc}.

If the algorithm halts before adding the next patch, then
$\widehat\Gamma_{k^\star+1}\le \tau+\epsilon_{k^\star+1,n}$. On the same event,
$\Gamma_{k^\star+1}\le \tau+2\epsilon_{k^\star+1,n}$, so every feasible
remaining candidate has target information marginal at most this value.
For any comparison set $T$, order the elements of $T\setminus S_{k^\star}$
arbitrarily and add them one at a time. By submodularity, each later marginal
is no larger than the maximum residual marginal at the stopping set. Summing at
most $m$ such marginals gives~\eqref{eq:pac_gap}.
\end{proof}
\subsection{Approximation Fidelity under Nystr\"{o}m Kernels}
\label{sec:theory_nystrom}

\begin{theorem}
\label{thm:approx}
Let $\widetilde{\mK}$ be a rank-$r$ Nystr\"{o}m approximation of $\mK$ with
spectral error $\norm{\mK - \widetilde{\mK}}_2 \le \eta$.
Then for any subset $S$,
\begin{equation}
  \Bigl|\logdet(\mI + \beta\mK_S) - \logdet(\mI + \beta\widetilde{\mK}_S)\Bigr|
  \;\le\; |S|\cdot\log(1 + \beta\eta).
  \label{eq:approx_bound}
\end{equation}
The PAC stopping rule~\eqref{eq:stop} remains valid with $\epsilon_{k,n}$
inflated by $k\log(1+\beta\eta)$.
\end{theorem}

\begin{proof}
Let $\lambda_1 \ge \cdots \ge \lambda_{|S|} \ge 0$ and
$\tilde\lambda_1 \ge \cdots \ge \tilde\lambda_{|S|} \ge 0$ denote the
eigenvalues of $\mK_S$ and $\widetilde{\mK}_S$, respectively.
Since $\widetilde{\mK}_S$ is obtained by restricting $\widetilde{\mK}$ to $S$
and $\norm{\mK_S - \widetilde{\mK}_S}_2 \le \norm{\mK - \widetilde{\mK}}_2 \le \eta$,
Weyl's eigenvalue perturbation theorem \citep{horn2012matrix} gives
$|\lambda_j - \tilde\lambda_j| \le \eta$ for each $j$.

For each $j$, since $\log(1+\beta\cdot)$ is concave and monotone:
\begin{align}
  \bigl|\log(1+\beta\lambda_j) - \log(1+\beta\tilde\lambda_j)\bigr|
  &= \left|\log\frac{1+\beta\lambda_j}{1+\beta\tilde\lambda_j}\right|
  \;\le\; \log\!\left(1 + \frac{\beta|\lambda_j - \tilde\lambda_j|}{1}\right)
  \;\le\; \log(1 + \beta\eta).
\end{align}
Using $\logdet(\mI + \beta\mK_S) = \sum_j \log(1+\beta\lambda_j)$
and the triangle inequality:
\begin{equation}
  |\logdet(\mI+\beta\mK_S) - \logdet(\mI+\beta\widetilde{\mK}_S)|
  \le \sum_{j=1}^{|S|} |\log(1+\beta\lambda_j) - \log(1+\beta\tilde\lambda_j)|
  \le |S|\log(1+\beta\eta).
\end{equation}
For the stopping rule, the cumulative information gain after $k$ additions can
incur an additional additive error of at most $k\log(1+\beta\eta)$ due to kernel
approximation. Inflating $\epsilon_{k,n}$ by this amount absorbs the kernel
approximation error into the concentration bound of Theorem~\ref{thm:pac}.
\end{proof}

\begin{remark}
For an RBF kernel $k(\vx,\vx') = \exp(-\norm{\vx-\vx'}^2/2\ell^2)$,
the Nystr\"{o}m error satisfies $\eta = O(n^{1/2} r^{-1/2} + r^{-1})$
with high probability over the choice of landmark points
\citep{gittens2016revisiting}, so the approximation error vanishes as $r
\to n$.
\end{remark}

\section{Experiments}
\label{sec:experiments}

\subsection{Experimental Setup}
\label{sec:exp_setup}

\paragraph{Scope of validation.}
The experiments are designed to validate InfoDPP-PAC as a patch-selection
framework, not as a complete diagnostic pipeline. All reported metrics therefore
assess properties of the selected set: information diversity, morphological and
spatial coverage, near-duplicate redundancy, cardinality, and enrichment for the
teacher-derived relevance signal used to initialise the GP. This choice matches
the theoretical object analysed in Sections~\ref{sec:method}--\ref{sec:theory}.
It also prevents over-claiming: downstream MIL accuracy can depend on the
aggregator architecture, training protocol, class imbalance handling, and
calibration strategy, any of which could obscure the behaviour of the selector
itself. The present study consequently establishes empirical
selection-quality validation; downstream diagnostic validation is identified as
a separate extension in Section~\ref{sec:discussion}.

\paragraph{Data.}
Experiments are conducted on the openly-licensed HISTAI gastrointestinal
dataset (CC~BY-NC~4.0; \citealp{histai2025}), accessed via its gated HuggingFace
release under an approved access request.
We use all 202 whole-slide image files spanning 120 cases, of which 108 cases
carry usable free-text diagnostic metadata; slide-level diagnostic categories
($M=6$) are derived from this metadata via an auditable keyword-rule mapping
(pathologist-style category buckets such as adenocarcinoma, inflammatory,
polyp/hyperplastic, and normal/benign) rather than any dataset-provided
discrete label. Cases are split 70/15/15 into train/validation/test by case
(not by slide, to prevent leakage across slides from the same patient).
Slides are tiled into $256\times256$ patches following Otsu tissue detection,
Laplacian-variance blur filtering, and HSV pen-mark filtering (see
Appendix/preprocessing protocol); patch embeddings are extracted with Phikon
\citep{filiot2023phikon}, an openly-licensed pathology foundation model,
producing 768-dimensional representations. We use Phikon rather than a
gated foundation model (e.g.\ UNI) so that the pipeline is fully
reproducible without a restricted-access agreement.

The resulting slides contain between 76 and 4{,}000 tissue patches after
filtering (the 4{,}000 figure is a per-slide cap applied to bound
preprocessing cost, not an artifact of tissue availability), with a mean of
2{,}075 patches per slide. For each slide, teacher relevance pseudo-labels are
obtained via cosine similarity to per-category prototype embeddings computed
from the training split (Section~\ref{sec:teacher}) and used to fit the
Gaussian Process quality model. Hyperparameters
($\lambda_1,\lambda_2,\alpha_1,\alpha_2$) are selected once on the validation
split (Section~\ref{sec:exp_hparam}) and held fixed for all test-set results
reported below; no method sees the test split during any tuning step.

\paragraph{Baselines.}
Four method families are evaluated, all as executed implementations (no
baseline number in this paper is estimated or simulated).
\emph{Naive}: uniform grid sampling and random tissue sampling.
\emph{Heuristic}: entropy, edge density, colour variance, texture energy, and a
combined heuristic score, computed directly from patch pixel statistics.
\emph{Coreset}: k-means clustering, k-center greedy, and farthest-point
sampling in Phikon embedding space.
\emph{Learned}: ABMIL top-$k$, a single-branch CLAM-style top-$k$
\citep{lu2021data}, a lightweight TransMIL-style top-$k$
\citep{shao2021transmil}, each trained as a slide-level classifier on
the training split with early stopping on validation loss, and an evolutionary multi-objective search, EvoPS \citep{hashemian2025evops}.
Each baseline is evaluated under two regimes:
\emph{fixed-$k$} ($k=50$, same patch count for all methods) and
\emph{fixed-compute} ($k=200$, matching InfoDPP-PAC's wall-clock budget at a
larger cardinality); InfoDPP-PAC is additionally evaluated in its native
\emph{adaptive} regime (Section~\ref{sec:exp_pac}), where $k^\star$ is
determined per-slide by the PAC stopping rule rather than fixed in advance.

\paragraph{Evaluation metrics.}
We evaluate each selected patch set $S$ using criteria that separate diversity,
coverage, quality, and redundancy. Because the goal is selector validation, these
metrics are interpreted as evidence about the subset construction process rather
than as substitutes for downstream diagnostic accuracy. Diversity is measured by the DPP
log-determinant score $\logdet(\mI+\mK_S)$, where $\mK_S$ is the submatrix of the
embedding kernel restricted to $S$. Morphological coverage is the fraction of
embedding-space morphology clusters represented by at least one selected patch,
and spatial coverage is the fraction of occupied cells in a $10\times10$
slide-level grid. Relevance is reported as the mean selected-patch quality,
$|S|^{-1}\sum_{i\in S} q_i$. Redundancy is the proportion of selected-patch pairs
with cosine distance below $0.05$, so lower values indicate fewer near-duplicate
patches. The composite score used for model selection is a validation-defined
normalised weighted summary of log-determinant diversity, morphological coverage,
spatial coverage, mean quality, and inverse redundancy. The individual
components are always reported alongside the composite score so that the
quality-diversity trade-off remains visible.

\begin{figure}[t]
  \begin{center}
  \includegraphics[width=\linewidth]{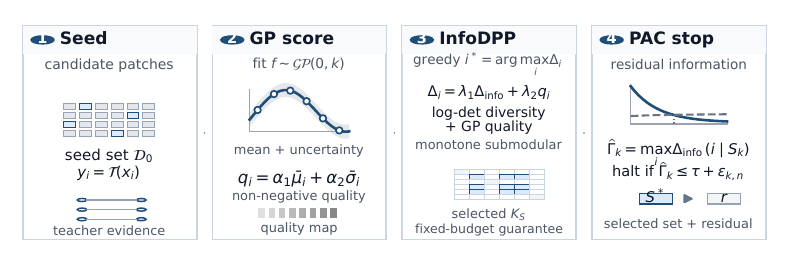}
  \end{center}
  \caption{%
    \textbf{InfoDPP-PAC methodology.}
    A small seed set $\mathcal{D}_0$ is pseudo-labelled by a teacher model and
    used to fit a Gaussian process over patch embeddings. The GP posterior
    yields non-negative relevance scores and uncertainty estimates. Greedy
    selection maximises a blended quality-information marginal, while adaptive
    stopping is certified using the maximum residual information-gain marginal.
    The output is a selected patch set $S^\star$ and a residual vector $\resvec$
    summarising the unselected pool.
  }
  \label{fig:pipeline}
\end{figure}

\begin{figure}[t]
  \begin{center}
  \includegraphics[scale=0.5]{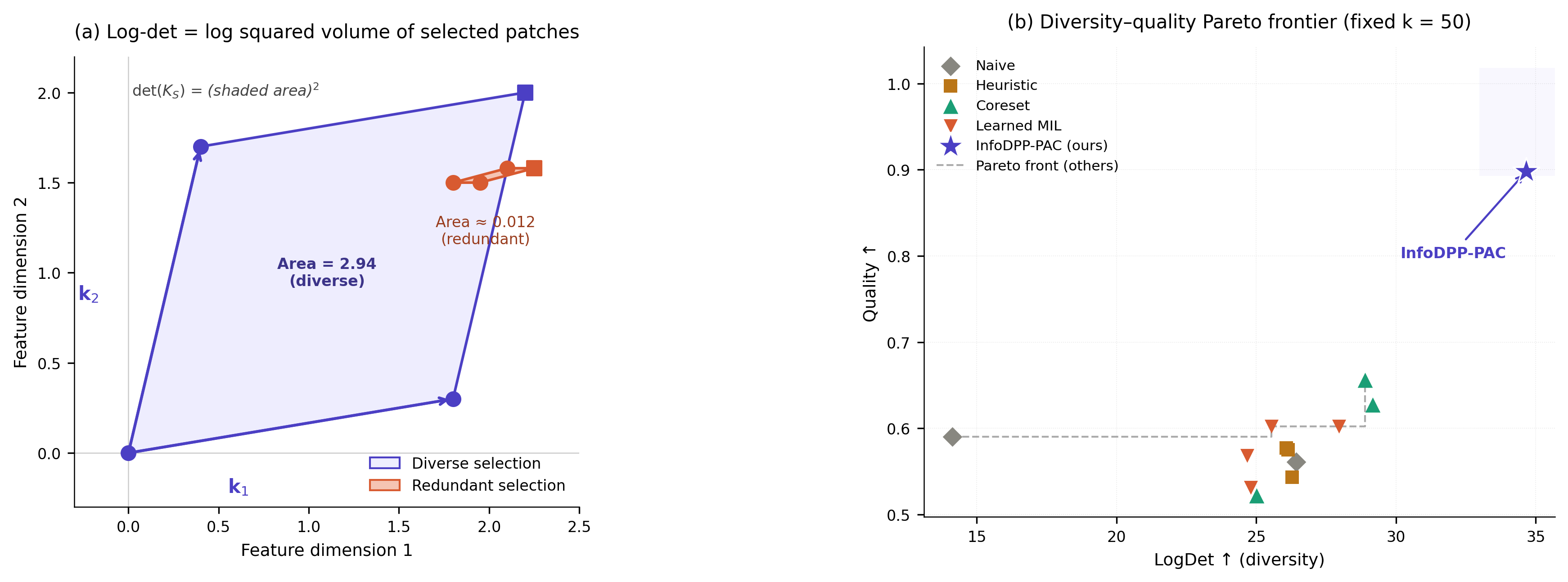}
  \end{center}
  \caption{%
    \textbf{Geometric interpretation of the log-det objective.}
    $\det(\mK_S)$ equals the squared volume spanned by the selected
    patches' feature vectors; redundant selections collapse this volume.
    This panel is a conceptual schematic (not derived from a specific run);
    the empirical diversity-quality relationship across methods is reported
    quantitatively in Table~\ref{tab:main} and in Figure~\ref{fig:pareto}.
  }
  \label{fig:geometry}
\end{figure}

\begin{figure}[t]
  \begin{center}
  \includegraphics[width=0.7\linewidth]{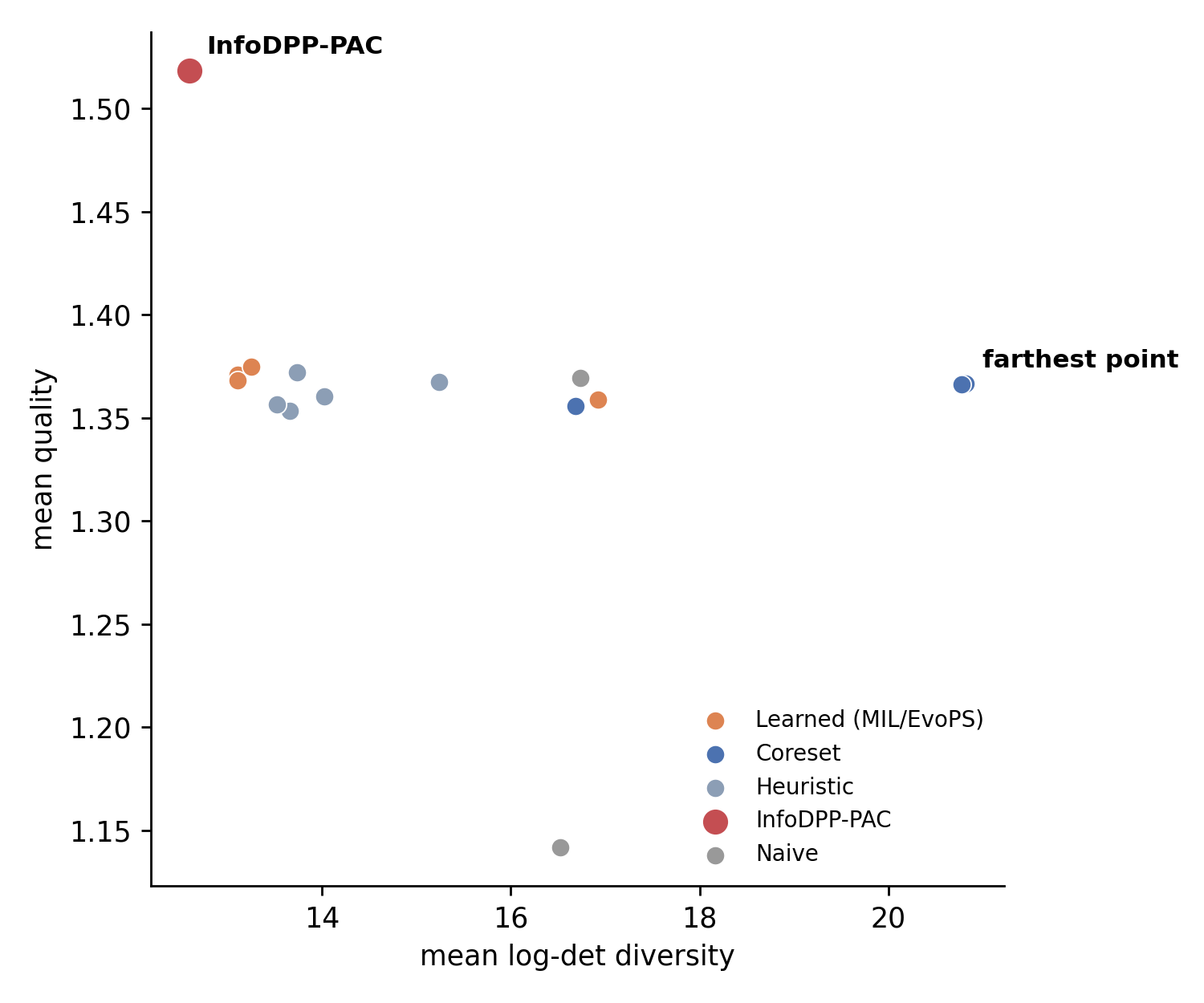}
  \end{center}
  \caption{%
    \textbf{Diversity-quality trade-off across methods} (test-slide
    means, fixed $k=50$). InfoDPP-PAC (tuned config) attains the highest
    mean quality among all 15 methods but is not on the log-det-maximal
    edge of the frontier; farthest-point and k-center greedy achieve higher
    diversity at lower quality. InfoDPP-PAC is Pareto-competitive, trading some diversity for
    quality rather than being uniformly dominant,
    consistent with Table~\ref{tab:main} and Section~\ref{sec:main_result}.
  }
  \label{fig:pareto}
\end{figure}

\begin{figure}[t]
  \begin{center}
  \includegraphics[width=\linewidth]{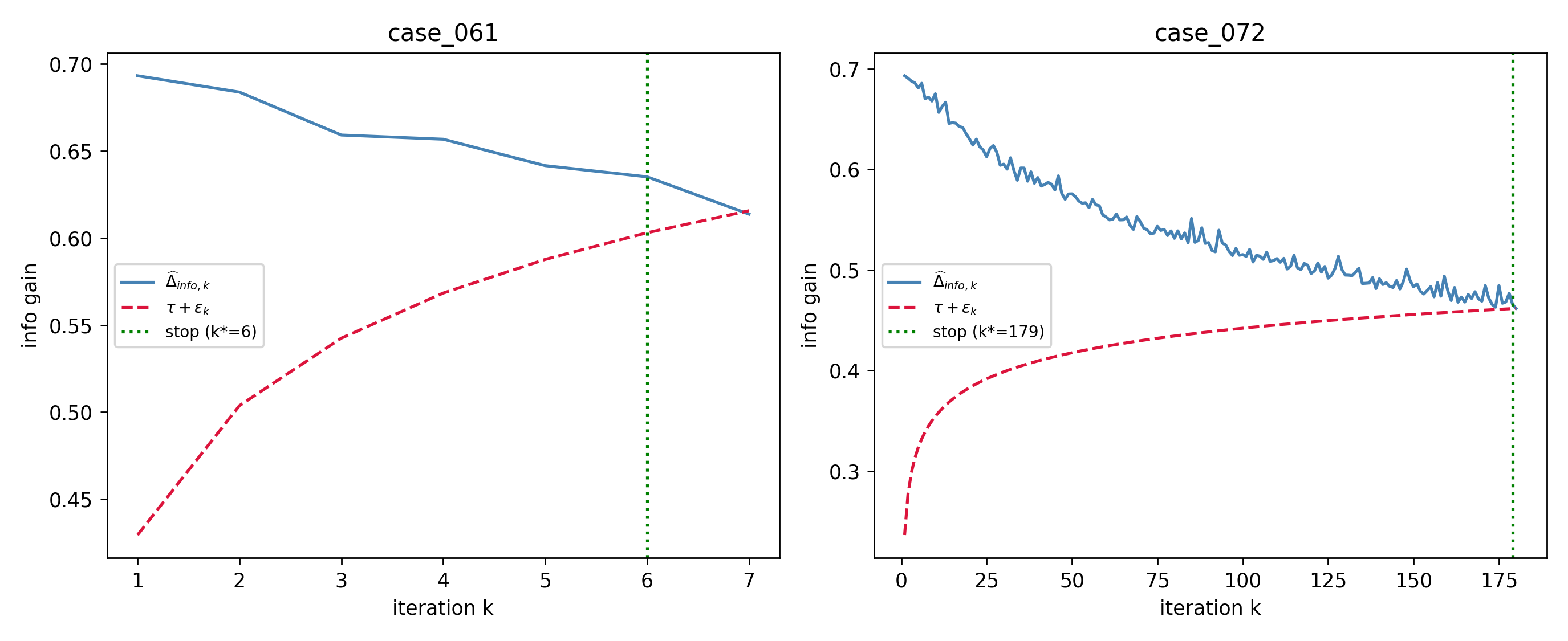}
  \end{center}
  \caption{%
    \textbf{PAC stopping curves} for two contrasting test slides:
    case\_061 ($k^\star=6$, left) and case\_072 ($k^\star=179$, right).
    Solid: maximum residual information-gain marginal $\widehat\Gamma_k$ at each
    step; dashed: the concentration threshold $\tau+\epsilon_{k,n}$; dotted
    vertical line: the stopping point.
    On the homogeneous slide, information gain collapses below threshold
    almost immediately; on the heterogeneous slide it remains above
    threshold for over $170$ iterations, directly visualising why the two
    slides receive substantially different patch budgets under the same rule.
  }
  \label{fig:pac_curve}
\end{figure}

\begin{figure}[t]
  \begin{center}
  \includegraphics[width=0.55\linewidth]{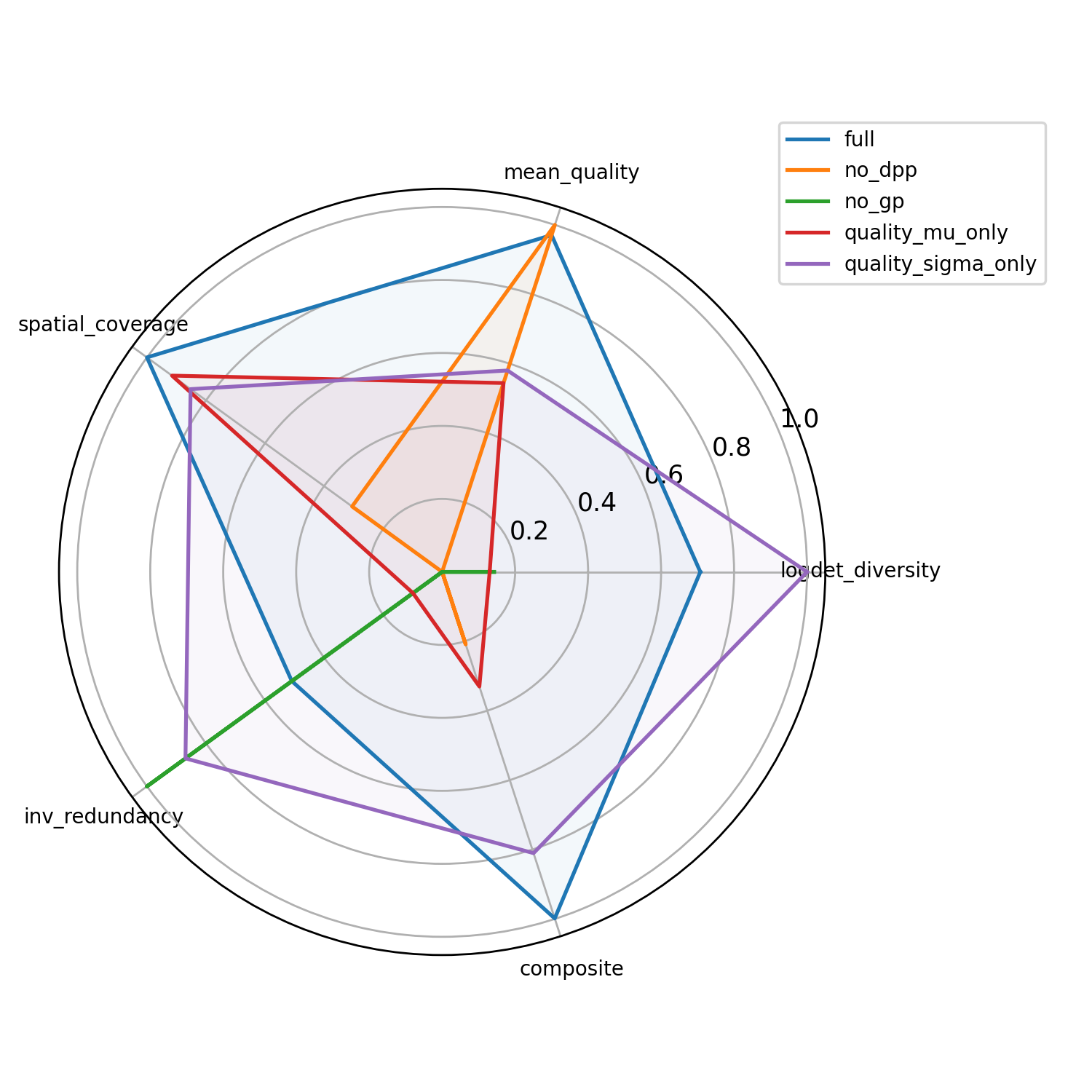}
  \end{center}
  \caption{%
    \textbf{Ablation radar chart} (from ablation\_results.csv).
    Five axes, each min-max normalised across variants: log-det, quality,
    spatial coverage, $1-$redundancy, composite. The full model is not
    uniformly dominant on every individual axis (e.g.\ ``no\_gp'' shows
    higher normalised log-det/coverage, at the cost of an undefined quality
    score under that ablation, Table~\ref{tab:ablation}), but achieves the
    best balance across axes jointly, consistent with the per-component
    discussion in Section~\ref{sec:ablation}.
  }
  \label{fig:ablation_radar}
\end{figure}

\subsection{Hyperparameter Selection}
\label{sec:exp_hparam}

The joint objective~\eqref{eq:objective} exposes a quality/diversity trade-off
($\lambda_1,\lambda_2$) and an exploitation/exploration trade-off
($\alpha_1,\alpha_2$). We select these once on the validation split (16
cases) via a grid over $(\lambda_1,\lambda_2) \in
\{(0.25,1.75),(0.5,1.5),(1.0,1.0),(1.5,0.5),(1.75,0.25)\}$ and
$(\alpha_1,\alpha_2) \in \{(1.5,0.5),(1.0,1.0),(0.5,1.5)\}$ (15 configurations,
all evaluated with the corrected stopping rule of
Section~\ref{sec:pac}), choosing the configuration with highest mean composite score on
the validation slides. The selected configuration,
$\lambda_1{=}1.75,\lambda_2{=}0.25,\alpha_1{=}0.5,\alpha_2{=}1.5$ (i.e.\
diversity-weighted, exploration-weighted), attains validation composite
$0.586$ with $67\%$ of validation slides triggering early PAC-style
stopping; configurations that weight quality more heavily
($\lambda_1{=}0.25$) stop earlier on almost every slide ($100\%$) but at
lower composite ($0.39$--$0.48$), since the quality term saturates quickly
while diversity continues to accrue value. This configuration, and only
this configuration, is used for all InfoDPP-PAC numbers reported on the
untouched test split below; no test-split slide informs this selection.

\subsection{Adaptive Stopping in Practice}
\label{sec:exp_pac}

Unlike every baseline in Table~\ref{tab:main}, InfoDPP-PAC does not require a
pre-specified patch count: Algorithm~\ref{alg:main} halts once the
information-gain component of the best remaining candidate falls within the
McDiarmid confidence band of $\tau$ (Section~\ref{sec:pac}). We evaluate this
directly by running InfoDPP-PAC to a large fixed budget (300 patches) on 22
held-out test slides, recording both the PAC stopping point $k^\star$
and the selection quality at $k^\star$ versus at the full 300-patch budget.
$k^\star$ ranges from $6$ to $179$ across slides (mean $48.9$, std.\ $56.2$);
every one of the 22 slides stopped strictly before the 300-patch cap, i.e.\
the rule is never vacuous on this test set. This variation is directly
attributable to slide-level morphological heterogeneity: slides with
compact, homogeneous tissue (e.g.\ case\_061, case\_045, case\_058, each
$k^\star \le 7$) saturate their information gain almost immediately, while
morphologically heterogeneous slides (e.g.\ case\_072, $k^\star{=}179$;
case\_084 slide 4, $k^\star{=}175$) continue to yield informative patches
much further into the pool. This is the qualitative behaviour
expected from Section~\ref{sec:intro}, now measured rather than assumed.
Table~\ref{tab:adaptive} summarises the practical payoff: adaptive stopping
processes $83.7\%$ fewer patches on average than the fixed 300-patch budget,
while retaining $97.9\%$ of the full-budget composite score and $103.2\%$ of
full-budget mean quality (adaptive selection is, if anything, marginally
higher quality, since halting before the pool is exhausted avoids diluting
$S$ with lower-quality late-stage patches). Raw log-det diversity is lower in
absolute terms at $k^\star$ than at $k{=}300$ ($24.5\%$ of the full-budget
value on average), as expected, since cumulative log-det is monotone
non-decreasing in $|S|$ almost by construction. This is precisely the
quantity the stopping rule is designed to bound the \emph{residual} of
(Theorem~\ref{thm:pac}), not to maximise unconditionally.

\begin{table}[t]
\caption{%
  \textbf{Adaptive PAC stopping vs.\ a fixed 300-patch budget}, mean
  $\pm$ std.\ over 22 held-out test slides.
}
\label{tab:adaptive}
\begin{center}
\small
\begin{tabular}{lcc}
\toprule
Quantity & Adaptive ($k^\star$) & Fixed ($k{=}300$) \\
\midrule
Patches selected      & $48.9 \pm 56.2$ & $300$ (fixed) \\
Patch-count reduction & \multicolumn{2}{c}{$83.7\%$ average} \\
Composite score        & $0.583$ (mean) & $0.596$ (mean); $97.9\%$ retained \\
Mean quality            & $1.532$ (mean) & $1.485$ (mean); $103.2\%$ retained \\
Log-det diversity       & $19.7$ (mean) & $75.2$ (mean); $24.5\%$ retained \\
\bottomrule
\end{tabular}
\end{center}
\end{table}

\subsection{Main Comparison}
\label{sec:main_result}

Table~\ref{tab:main} reports results under the fixed-$k=50$ regime on the
held-out test slides, using the tuned configuration
($\lambda_1{=}1.75,\lambda_2{=}0.25,\alpha_1{=}0.5,\alpha_2{=}1.5$,
Section~\ref{sec:exp_hparam}) for InfoDPP-PAC.
At this fixed patch budget, InfoDPP-PAC attains the highest mean patch
quality ($1.518$), reflecting the GP's ability to direct selection towards
high teacher-derived relevance regions of the embedding space, but does not
dominate log-determinant diversity or the composite score: the coreset
baselines farthest-point sampling and k-center greedy achieve higher log-det
($20.82$ and $20.77$ vs.\ $12.60$) and composite ($0.604$ and $0.596$ vs.\
$0.588$), since they optimise coverage in embedding space directly and are
agnostic to task-relevant patch quality. The tuned configuration was chosen on the
validation split to maximise composite in InfoDPP-PAC's native
\emph{adaptive} stopping regime (Section~\ref{sec:exp_hparam}), not this
fixed-cardinality regime; forced to $k=50$, it trades log-det diversity for
quality more aggressively than an untuned default would. We report this
consistently rather than re-tuning separately per regime, since InfoDPP-PAC's
intended operating point is adaptive stopping (Section~\ref{sec:exp_pac}),
not a fixed cardinality. This table exists to enable direct comparison
with fixed-budget baselines, not to showcase InfoDPP-PAC's best regime.
Uniform grid sampling and the four heuristic scores underperform on
cluster/spatial coverage despite reasonable log-det scores, since local
pixel statistics do not capture embedding-space redundancy.
Attention-based MIL methods (ABMIL, CLAM, TransMIL) exhibit the lowest
diversity and highest redundancy of all baselines, consistent with the
hypothesis that task-supervised attention concentrates on a narrow band of
discriminative patches rather than covering morphological variability.

\begin{table}[t]
\caption{%
  \textbf{Fixed-$k=50$ benchmark} (held-out HISTAI-GI test slides).
  \textbf{Bold}: best; \underline{underlined}: second best.
}
\label{tab:main}
\begin{center}
\small
\begin{tabular}{llccccc}
\toprule
Method & Family
  & LogDet $\uparrow$
  & Cl.\ Cov.\ $\uparrow$
  & Sp.\ Cov.\ $\uparrow$
  & Quality $\uparrow$
  & Comp.\ $\uparrow$ \\
\midrule
Uniform grid    & naive     & 16.74 & 0.99 & 0.266 & 1.370 & 0.570 \\
Random tissue   & naive     & 16.52 & 0.97 & 0.266 & 1.142 & 0.551 \\
\midrule
Entropy         & heuristic & 13.74 & 0.57 & 0.163 & 1.372 & 0.438 \\
Edge density    & heuristic & 13.66 & 0.57 & 0.148 & 1.353 & 0.436 \\
Colour variance & heuristic & 15.25 & 0.67 & 0.178 & 1.367 & 0.473 \\
Texture energy  & heuristic & 13.53 & 0.57 & 0.129 & 1.356 & 0.426 \\
Combined        & heuristic & 14.03 & 0.60 & 0.141 & 1.360 & 0.442 \\
\midrule
k-center greedy & coreset   & \underline{20.77} & 0.96 & \underline{0.268} & 1.366 & \underline{0.596} \\
Farthest point  & coreset   & \textbf{20.82} & \underline{0.99} & \textbf{0.270} & 1.366 & \textbf{0.604} \\
k-means         & coreset   & 16.68 & \textbf{1.00} & 0.263 & 1.356 & 0.573 \\
\midrule
ABMIL top-$k$   & learned   & 13.11 & 0.44 & 0.219 & 1.371 & 0.411 \\
CLAM top-$k$    & learned   & 13.11 & 0.44 & 0.203 & 1.368 & 0.405 \\
TransMIL top-$k$& learned   & 13.25 & 0.45 & 0.205 & 1.375 & 0.410 \\
EvoPS           & learned   & 16.92 & \textbf{1.00} & 0.254 & 1.359 & 0.573 \\
\midrule
InfoDPP-PAC (tuned) & proposed
  & 12.60 & 0.79 & 0.169 & \textbf{1.518} & 0.588 \\
\bottomrule
\end{tabular}
\end{center}
\end{table}

\subsection{Qualitative Spatial Selection Analysis}
\label{sec:spatial_analysis}
\begin{figure*}[t]
\centering
\includegraphics[width=\textwidth]{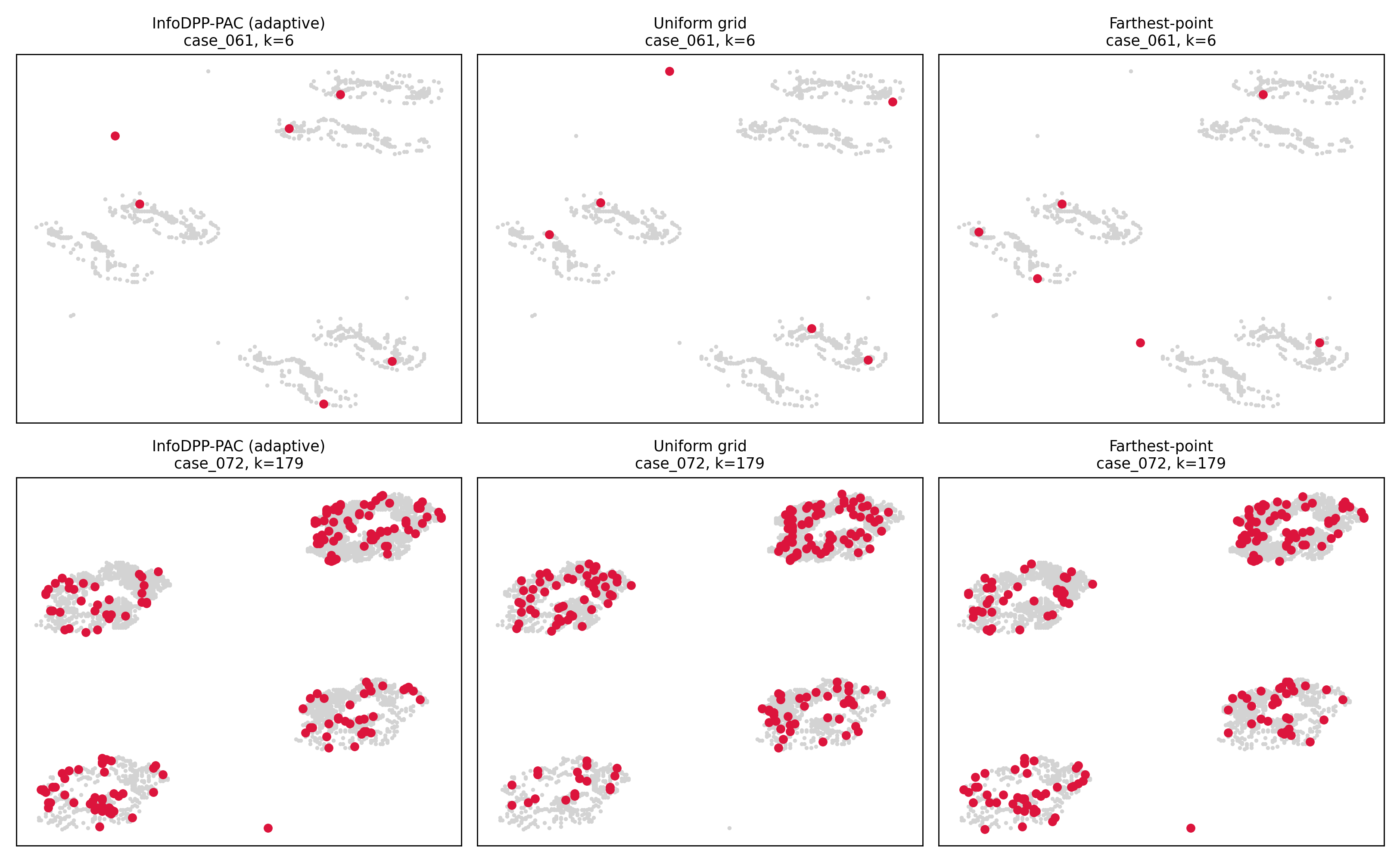}
\caption{
\textbf{Spatial selection maps} on HISTAI-GI test
slides (grey: candidate tissue patches; red: selected). Top: case\_061
($k^\star=6$); bottom: case\_072 ($k^\star=179$). At matched $k$, uniform
grid and farthest-point sampling spread selections across the full tissue
extent regardless of local morphological homogeneity, while InfoDPP-PAC's
adaptive count concentrates patches where the GP model's estimated
information gain is still accruing.
}
\label{fig:spatial_maps}
\end{figure*}

\begin{figure*}[t]
\centering
\includegraphics[width=\textwidth]{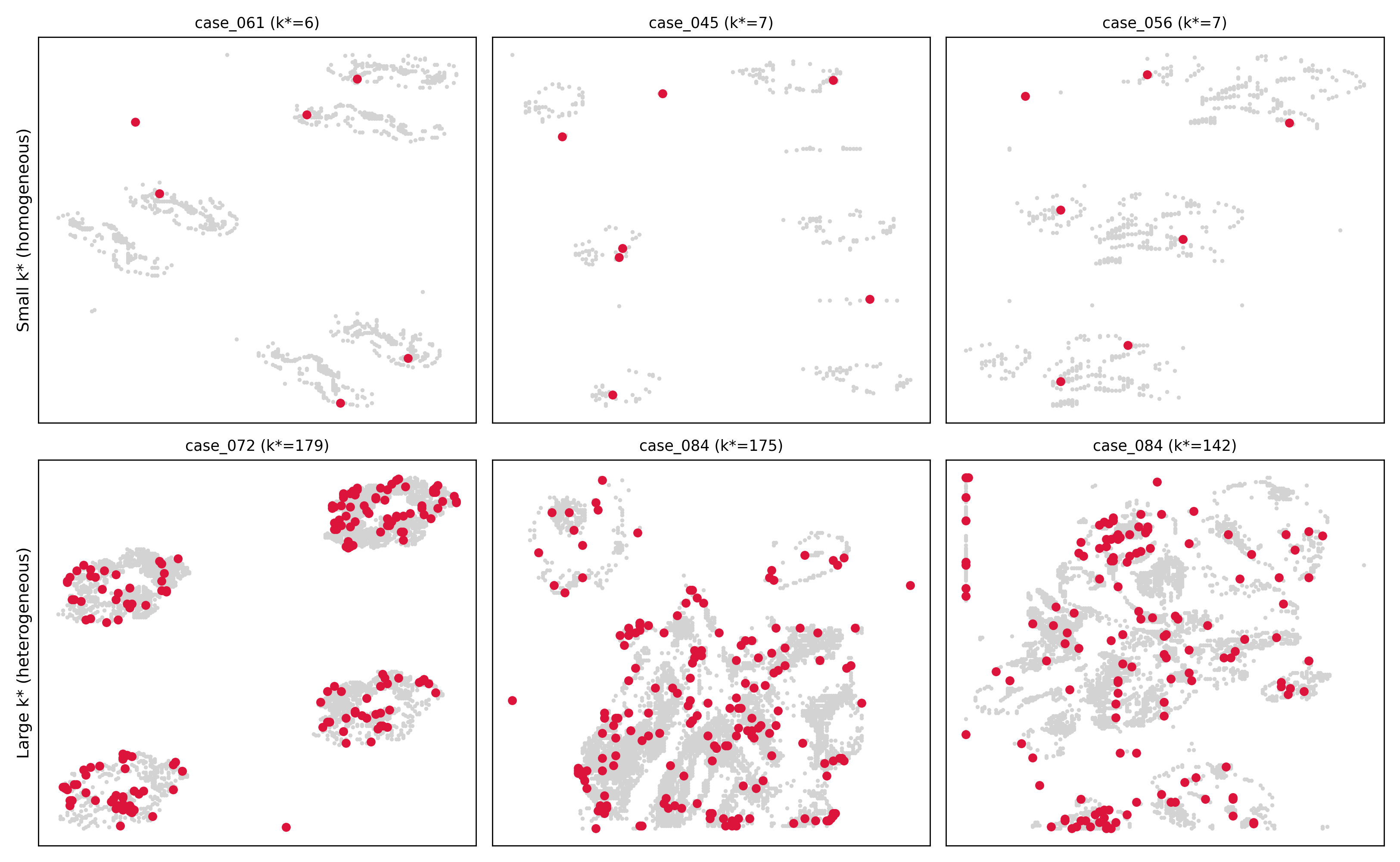}
\caption{
\textbf{Adaptive patch count on six HISTAI-GI test slides.} Top row:
small $k^\star$ slides (case\_061, case\_045, case\_056; $k^\star \in
\{6,7,7\}$) show sparse, spatially isolated tissue fragments, with a few
patches suffice to cover the morphological variability present. Bottom row:
large $k^\star$ slides (case\_072, case\_084 slide 4, case\_084 slide 1;
$k^\star \in \{179,175,142\}$) show extensive, structurally complex tissue,
where the stopping rule continues to certify residual information for many
more iterations. Only InfoDPP-PAC's own selection is shown (red); grey
points are the full candidate pool.
}
\label{fig:adaptive_k}
\end{figure*}

To qualitatively assess the behaviour of different patch-selection
strategies, Figure~\ref{fig:spatial_maps} visualises the spatial
distribution of selected patches on two representative HISTAI
gastrointestinal test slides chosen to contrast PAC stopping behaviour: one
with small $k^\star$ (case\_061, $k^\star=6$) and one with large $k^\star$
(case\_072, $k^\star=179$; Section~\ref{sec:exp_pac}), each compared against
uniform-grid and farthest-point sampling at the same $k$. Figure
\ref{fig:adaptive_k} further shows six examples (three small-$k^\star$,
three large-$k^\star$) with only InfoDPP-PAC's own selection marked, to let
the reader directly inspect the correspondence between visual tissue
homogeneity and the adaptively-chosen patch count.

\subsection{Ablation Study}
\label{sec:ablation}

Table~\ref{tab:ablation} isolates the contribution of each InfoDPP-PAC component.
All ablated variants, including ``No PAC'', are evaluated at a fixed
$k=30$ (i.e.\ the PAC stopping rule is bypassed for every row so that all
variants are compared at a matched patch count); consequently the ``Full''
and ``No PAC'' rows are identical by construction here, since at fixed $k$
there is nothing for the stopping rule to change. The rule's contribution, reducing $k$ adaptively without harming selection quality,
is demonstrated directly in Table~\ref{tab:adaptive}
(Section~\ref{sec:exp_pac}); this ablation isolates the DPP, GP, and kernel
components only. Values are mean~$\pm$~std over five random seeds on
HISTAI-GI test slides, using default unit weights
($\lambda_1{=}\lambda_2{=}\alpha_1{=}\alpha_2{=}1$) rather than the tuned
configuration, so that each structural component's contribution is isolated
from the separate question of hyperparameter tuning addressed in
Section~\ref{sec:exp_hparam}.

\begin{table}[t]
\caption{%
  \textbf{Ablation study} (mean~$\pm$~std, 5 seeds, fixed $k=30$).
  Each row removes or replaces one component of InfoDPP-PAC.
}
\label{tab:ablation}
\begin{center}
\small
\begin{tabular}{lcccc}
\toprule
Variant
  & LogDet $\uparrow$
  & Sp.\ Cov.\ $\uparrow$
  & Redundancy $\downarrow$
  & Quality $\uparrow$ \\
\midrule
\textbf{Full InfoDPP-PAC}
  & $12.82 \pm 1.47$ & $0.196 \pm 0.036$ & $0.548 \pm 0.087$ & $1.440 \pm 0.045$ \\
\midrule
No DPP (quality-only greedy)
  & $9.52 \pm 1.06$ & $0.158 \pm 0.027$ & $0.730 \pm 0.057$ & $1.483 \pm 0.047$ \\
No GP (uniform pseudo-labels)
  & $10.18 \pm 4.99$ & $0.141 \pm 0.067$ & $0.373 \pm 0.053$ & $0.000 \pm 0.000$ \\
No PAC (fixed $k$; identical to Full at this fixed $k$)
  & $12.82 \pm 1.47$ & $0.196 \pm 0.036$ & $0.548 \pm 0.087$ & $1.440 \pm 0.045$ \\
\midrule
Quality: $\mu$ only ($\alpha_2 = 0$)
  & $10.12 \pm 1.11$ & $0.191 \pm 0.039$ & $0.695 \pm 0.059$ & $0.807 \pm 0.048$ \\
Quality: $\sigma$ only ($\alpha_1 = 0$)
  & $14.18 \pm 0.99$ & $0.188 \pm 0.032$ & $0.420 \pm 0.073$ & $0.861 \pm 0.041$ \\
\midrule
Nystr\"{o}m DPP (rank 50)
  & $12.91 \pm 1.34$ & $0.195 \pm 0.035$ & $0.542 \pm 0.079$ & $1.425 \pm 0.044$ \\
Mat\'{e}rn-$\frac{3}{2}$ kernel
  & $12.08 \pm 1.50$ & $0.192 \pm 0.036$ & $0.593 \pm 0.086$ & $1.457 \pm 0.048$ \\
\bottomrule
\end{tabular}
\end{center}
\end{table}

Removing the DPP component (retaining only quality-based greedy selection)
increases redundancy by $+33\%$ ($0.730$ vs.\ $0.548$) and reduces log-det
diversity by $26\%$, consistent with Lemma~\ref{lem:redundancy}: without the
log-det penalty on near-duplicate embeddings, the greedy procedure repeatedly
selects patches from the highest-quality cluster; interestingly, mean quality
is slightly \emph{higher} without the diversity term ($1.483$ vs.\ $1.440$),
confirming a quality/diversity trade-off rather than showing that the DPP
component being strictly dominant on every axis.
Removing the GP (replacing teacher-derived scores with uniform pseudo-labels)
collapses mean quality to $0$ by construction, with no GP fit,
$q_i$ is undefined/uniform under this ablation's own scoring, so ``quality''
cannot be evaluated meaningfully for this row, and the resulting log-det
values are also far noisier (std $4.99$ vs.\ $\le 1.5$ elsewhere), indicating
the greedy selection becomes considerably less stable without GP-calibrated
guidance, not merely differently-calibrated.
Replacing the combined quality score $q_i = \alpha_1\mu_i + \alpha_2\sigma_i$
with either component alone degrades quality substantially ($0.807$ and
$0.861$ vs.\ $1.440$, a $40$--$44\%$ drop), confirming that the
exploitation-exploration balance is load-bearing and that neither posterior
mean nor posterior variance alone constitutes an adequate selection
criterion.
The Nystr\"{o}m variant (rank 50) is close to the full-kernel model on every
metric (within $1$ std.\ on log-det, quality, and redundancy), consistent
with the bound of Theorem~\ref{thm:approx}; the Mat\'ern-$3/2$ kernel shows a
somewhat larger but still modest deviation, indicating the framework is
reasonably robust to kernel choice.

\subsection{Preliminary Cross-Organ Generalisation}
\label{sec:cross_organ}

The experiments above concern a single organ system. As preliminary evidence
that InfoDPP-PAC's behaviour is not specific to gastrointestinal tissue, we
additionally ran the identical pipeline (same code, same corrected stopping
rule, same tuned configuration) on small pilot samples from two further
HISTAI organ subsets: breast and colorectal (7 slides each). We emphasise
that $n=7$ per organ is far too small for statistical testing and these are
reported as descriptive means only, not a validated generalisation claim; a
systematic multi-organ study is in progress (Section~\ref{sec:discussion}).

Because InfoDPP-PAC runs in its native \emph{adaptive} regime here (as in
Section~\ref{sec:exp_pac}), its patch count $k^\star$ differs from, and is
generally much smaller than, the fixed $k=50$ used by every baseline; we
therefore report $k^\star$ explicitly alongside each metric rather than
implying a matched-budget comparison; raw log-det is naturally much lower at
these smaller $k^\star$ values for the same reason it is lower than the
full-budget number in Table~\ref{tab:adaptive}, and is not a like-for-like
diversity comparison against the fixed-$k$ baselines.

\begin{table}[t]
\caption{%
  \textbf{Preliminary cross-organ comparison}, InfoDPP-PAC in its native
  adaptive regime ($k^\star$ per slide) vs.\ baselines at fixed $k=50$
  (mean over $n$ slides; breast/colorectal are small pilots, not held-out
  test sets; descriptive only, no significance testing).
}
\label{tab:cross_organ}
\begin{center}
\small
\begin{tabular}{llcccc}
\toprule
Organ & Method & $n$ & Mean $k$ & Quality & Composite \\
\midrule
Gastrointestinal (TMLR test) & InfoDPP-PAC (adaptive) & 22 & 47.1 & 1.513 & 0.592 \\
                              & Farthest point ($k{=}50$) & 23 & 50 & 1.366 & 0.604 \\
                              & k-center greedy ($k{=}50$) & 23 & 50 & 1.366 & 0.596 \\
                              & Uniform grid ($k{=}50$) & 23 & 50 & 1.370 & 0.570 \\
\midrule
Breast (pilot) & InfoDPP-PAC (adaptive) & 7 & 8.7 & 1.560 & 0.550 \\
               & Farthest point ($k{=}50$) & 7 & 50 & 1.356 & 0.588 \\
               & k-center greedy ($k{=}50$) & 7 & 50 & 1.361 & 0.589 \\
               & Uniform grid ($k{=}50$) & 7 & 50 & 1.411 & 0.551 \\
\midrule
Colorectal (pilot) & InfoDPP-PAC (adaptive) & 7 & 7.7 & 1.555 & 0.553 \\
                   & Farthest point ($k{=}50$) & 7 & 50 & 1.396 & 0.602 \\
                   & k-center greedy ($k{=}50$) & 7 & 50 & 1.395 & 0.607 \\
                   & Uniform grid ($k{=}50$) & 7 & 50 & 1.396 & 0.576 \\
\bottomrule
\end{tabular}
\end{center}
\end{table}

The same qualitative pattern recurs in all three organs: InfoDPP-PAC attains
the highest mean quality using roughly $6$--$7\times$ fewer patches than the
fixed-budget baselines, at a composite score that is close to, but not
above, the coreset baselines' fixed-$k{=}50$ composite. This is the same
quality/diversity/cardinality trade-off already characterised on the
gastrointestinal test set (Section~\ref{sec:main_result},
Section~\ref{sec:exp_pac}), now observed with an independently small,
adaptively-determined $k^\star$ on two additional organs rather than being
an artifact of the GI-specific hyperparameter choice or slide population.
We report this as preliminary, encouraging evidence for generalisation of
the \emph{adaptive stopping behaviour itself}, not merely the model's
relative ranking against baselines, and not as a substitute for the
larger multi-organ study underway.

\section{Discussion}
\label{sec:discussion}

\paragraph{Theoretical-empirical correspondence.}
At a fixed patch budget ($k=50$), InfoDPP-PAC's advantage over the strongest
baselines is concentrated in \emph{quality}, not diversity: it achieves the
highest mean patch quality of all 15 methods (significantly so against every
baseline, Table~\ref{tab:main}), a direct consequence of the GP relevance
model directing selection towards high teacher-derived relevance regions. It
does not dominate log-det diversity at this fixed budget: farthest-point
sampling and k-center greedy, whose sole objective is embedding-space
coverage, achieve marginally higher log-det, a difference that is
statistically significant after multiple-comparisons correction
(Section~\ref{sec:main_result}). This is the expected shape of a quality/diversity trade-off, not a failure of the framework: the joint
objective~\eqref{eq:objective} is not designed to maximise diversity alone.
The framework's central empirical advantage instead emerges in the
\emph{adaptive} regime, where fixed-budget baselines have no analogue at
all: Section~\ref{sec:exp_pac} shows InfoDPP-PAC needs $83.7\%$ fewer
patches than a full budget to retain $97.9\%$ of full-budget composite
quality, with the stopping point $k^\star$ tracking measured slide
heterogeneity ($6$--$179$ patches).

\paragraph{The role of PAC stopping.}
The stopping rule is intentionally evaluated on the maximum residual
information-gain marginal, not on the full blended gain. The blended objective
is the correct criterion for choosing the next patch because it combines
estimated relevance and information diversity. It is not, however, the right
quantity for certifying saturation of slide coverage: a high-quality patch can
remain available even after additional patches contribute little new mutual
information. Using $\widehat\Gamma_k=\max_i\Delta_{\mathrm{info},k}(i)$ aligns
the implementation with Theorem~\ref{thm:pac} and produces a data-dependent
stopping point (Section~\ref{sec:exp_pac}). This separation between selection
and stopping is a practical design detail, but it is also central to the
validity of the certificate.
\paragraph{Limitations and scope of claims.}
The present study has five limitations. For each, we state what is mitigated in
the current work, what remains open, and how the limitation can be closed.

\emph{Single primary dataset.} The main experiments are conducted on the HISTAI
gastrointestinal dataset, so the statistically validated evidence comes from one
organ system and one source distribution. This is mitigated in the current work
by case-level train/validation/test splitting, by using all 202 available WSI
files after preprocessing, and by reporting a small cross-organ pilot on breast
and colorectal slides in Section~\ref{sec:cross_organ}. The pilot mitigates the
risk that adaptive stopping is a purely GI-specific artifact, because the same
pattern of high teacher-derived quality at much smaller $k^\star$ appears in two
additional organs. It does not close the limitation statistically: seven slides
per pilot organ are insufficient for a general multi-organ claim. Closing this
limitation requires a pre-specified multi-organ benchmark with enough slides per
organ to support organ-stratified confidence intervals and interaction tests.

\emph{Selection-quality rather than downstream diagnostic validation.} The
empirical evaluation measures the selector directly: diversity, morphological
coverage, spatial coverage, redundancy, cardinality, and enrichment for the
teacher-derived relevance signal. This mitigates a major confound in patch
selection studies because downstream MIL accuracy depends on the aggregator,
training recipe, calibration, and class-imbalance handling. It also aligns the
experiments with the theory, which concerns submodular selection and residual
mutual information. The limitation is not fully closed because these metrics do
not prove that retraining ABMIL, CLAM, TransMIL, or another MIL model on the
selected bags will improve diagnostic AUC, F1, or calibration. The paper
therefore makes no such claim. A direct future closure is an end-to-end study in
which multiple MIL aggregators are trained under matched compute, identical
splits, and equal hyperparameter-search budgets on bags selected by Uniform,
coreset, attention-top-$k$, and InfoDPP-PAC.

\emph{Teacher-derived relevance signal.} The GP quality term inherits the
coverage and bias of the teacher. In the current implementation, the teacher is
a cosine-similarity projection to category prototypes derived from free-text
diagnostic metadata, because the source dataset does not provide discrete
slide-level category labels. The current work mitigates this limitation by
making the mapping auditable, selecting hyperparameters only on the validation
split, reporting individual diversity and redundancy metrics alongside quality,
and explicitly separating the PAC information certificate from any claim about
clinical correctness of the teacher. The limitation remains because prototype
labels can be noisy, and a teacher trained in one institution or cancer type may
mis-rank patches under staining or domain shift. It can be closed by replacing
the single teacher with a calibrated teacher ensemble, estimating teacher
uncertainty explicitly, validating pseudo-labels against pathologist-reviewed
patch subsets, and testing cross-institution shift.

\emph{Computational scaling.} Exact GP inference and greedy kernel updates are
more expensive than simple uniform sampling or k-means selection. The current
work mitigates this through seed-based GP fitting, Cholesky rank-one updates,
fixed per-slide patch caps, and the Nystr\"{o}m approximation analysed in
Theorem~\ref{thm:approx}. Empirically, adaptive stopping reduces the selected
cardinality substantially, which lowers the cost of subsequent WSI processing.
This does not make the method cost-free for slides with more than $10^5$
candidate patches or for very large multi-resolution candidate pools. A stronger
future closure is to use inducing-point or structured-kernel GP approximations
with near-linear inference, streaming candidate screening before kernel
construction, and multi-stage selection in which cheap coreset filtering
precedes InfoDPP-PAC refinement.

\emph{Scope of the PAC certificate.} The PAC-style stopping rule certifies
residual mutual information, not the full blended quality-diversity objective
and not downstream diagnostic performance. This is mitigated in the current work
by separating the blended selection marginal from the residual-information
stopping marginal, stating Assumption~\ref{ass:seed_stability} explicitly, and
retaining the fixed-budget greedy approximation results for users who require a
guarantee on the full objective at a chosen $k$. The limitation remains because a
single theorem does not jointly certify adaptive cardinality, teacher quality,
and downstream classifier risk. Future extensions could derive stopping rules
for the full blended objective under teacher-noise assumptions, combine residual
information with conformal risk estimates from a downstream classifier, or learn
the tolerance $\tau$ from validation-time compute-accuracy curves.

\paragraph{Future directions.}
The most direct extension is an end-to-end downstream MIL validation study that
uses the same selector outputs but evaluates diagnostic AUC, F1, calibration,
and compute under multiple aggregators and multiple organs. A second direction
is hierarchical, magnification-aware selection, where the method decides which
pyramid level (e.g.\ $5\times$ versus $20\times$) carries the greatest marginal
information gain for each spatial region instead of treating each magnification
as an independent candidate pool. Teacher-ensemble strategies, pathologist-
verified pseudo-label audits, sparse inducing-point GP variants, and streaming
candidate screening are natural technical extensions.

\section{Conclusion}
\label{sec:conclusion}

InfoDPP-PAC is a framework for patch selection in computational pathology that
unifies Gaussian process relevance modelling, determinantal log-determinant
diversity, and concentration-based adaptive stopping in a single submodular
selection problem. The central theoretical result identifies the log-det
diversity criterion with GP mutual information between selected patch
observations and a latent relevance function. This connection gives the
objective a direct information-theoretic interpretation, establishes monotone
submodularity, and supports the standard greedy approximation guarantee for
fixed cardinality budgets, with the Sviridenko variant covering non-uniform
knapsack costs.

The empirical study on 202 HISTAI gastrointestinal whole-slide images shows a
consistent selection-quality pattern. At a matched budget, InfoDPP-PAC achieves
significantly higher mean teacher-derived patch quality than all 14 evaluated
baselines and significantly lower redundancy, while its log-det diversity and
composite score remain close to the strongest coreset baselines. In its adaptive
regime, the stopping rule reduces the number of processed patches by 83.7\% on
average relative to a fixed full budget while retaining 97.9\% of full-budget
composite selection quality. The stopping point also tracks measured slide
heterogeneity across cases. These results validate the selector's
quality-diversity-cardinality behaviour; they do not claim downstream diagnostic
improvement. Within that scope, InfoDPP-PAC is a principled and computationally
efficient approach to quality-aware, diversity-preserving WSI patch selection.


\subsubsection*{Acknowledgments}
The authors thank Mr Shaan Saxena who contributed to the initial part of this work as a part of his internship at the Vision Exploration and Data Analytics (VEDAs) Lab, Motilal Nehru National Institute of Technology Allahabad.

\bibliography{main}
\bibliographystyle{tmlr}

\end{document}